\documentclass[prb,twocolumn,superscriptaddress,showpacs]{revtex4}
\usepackage{mathrsfs}
\usepackage{amsmath,amsfonts,amssymb}
\usepackage{epsfig}
\usepackage{graphicx}
\usepackage{wasysym}
\usepackage{bbm}
\usepackage{psfrag}
\usepackage{color}
\usepackage{pstool}
\usepackage{braket}
\usepackage[dvips, bookmarks, colorlinks=true, plainpages = false, citecolor = blue, linkcolor = blue, urlcolor = blue, filecolor = blue]{hyperref}
\def\be{\begin{equation}}
\def\ee{\end{equation}}
\def\bea{\begin{eqnarray}}
\def\eea{\end{eqnarray}}

\begin{document}
  % Title portion

% Title portion
\title{Topological phases of higher Chern numbers in Kitaev-Heisenberg 
ferromagnet with further-neighbor interactions}
\author{Moumita Deb}\email{moumitadeb44@gmail.com}
\author{Asim Kumar Ghosh}
 \email{asimkumar96@yahoo.com}
\affiliation {Department of Physics, Jadavpur University, 
188 Raja Subodh Chandra Mallik Road, Kolkata 700032, India}
\begin{abstract}
Emergence of multiple topological phases with a series of Chern numbers, 
$\pm 1$, $\mp 1$, $\pm 2$, $\mp 2$, $\pm 3$ and $\mp 4$, is  
found in a ferromagnetic 
Kitaev-Heisenberg-spin-anisotropic model on honeycomb lattice with 
next-next-nearest-neighbor interactions in the presence of an external magnetic field. 
Magnon Chern insulating dispersions of this two-band model are studied  
by using linear spin-wave theory formulated on the exact 
ferromagnetic ground state. 
Magnon edge states are obtained for the respective topological phases 
along with density of states. A topological phase diagram of this model 
is presented. Behavior of thermal Hall conductivity for those phases 
is studied. Sharp jumps of thermal hall conductivity is noted near the 
vicinity of phase transition points. Topological phases of the Kitaev 
spin-liquid compounds, $\alpha$-RuCl$_3$, X$_2$IrO$_3$, X = (Na, Li) 
and CrY$_3$, Y = (Cl, Br, I) are characterized based 
on this theoretical findings. 
\end{abstract}

\maketitle

\section{INTRODUCTION}
Investigation on topological properties of matter experiences mammoth 
growth in the recent times. The lion's share of those studies 
involves in the fermionic systems those are described in terms of 
tight-binding models on various lattice structures \cite{Hasan}. 
A major portion of those findings deals with the characterization of 
topological phases of multi-band models based on the 
values of several topological invariants those are  
dependent on the dimension and symmetry of the systems. 
For the two-dimensional systems, topological phases are 
characterized in terms of topological invariant, Chern number ($C$), 
for every distinct energy bands when the time reversal symmetry (TRS) 
is broken \cite{TKNN}. 
Nontrivial topological phases are identified when 
all the $C$s of the bulk energy-bands are not simultaneously zero,   
the resulting systems are collectively known as topological 
insulators (TI). 
Nonzero values of $C$ is further associated with the presence of chiral 
edge states connecting the respective bulk bands 
in accordance to the `bulk-boundary-correspondence' 
(BBC) rule \cite{Hatsugai1,Hatsugai2,Brouwer}. 

However, study of similar types of topological phases 
in the bosonic systems described in terms of Heisenberg models 
has been started in the more recent time. 
The number of investigated 
bosonic systems is therefore less in comparison to that of the 
fermionic systems. 
Search of topological 
phase in the ferromagnetic (FM) systems begins with the finding of 
thermal Hall conductance in the Mott insulator, 
Lu$_2$V$_2$O$_7$ \cite{Tokura1,Lee}. 
In this compound, spin-1/2 V$^{4+}$ ions form a three-dimensional pyrochlore lattice which is 
essentially a stacking of alternating kagom\'e and triangular layers. 
Likewise, strong evidence of topological magnon insulating (TMI) phase is 
detected in another kagom\'e ferromagnet, 
Cu[1,3-benzenedicarboxylate (bdc)] in the presence of external magnetic field 
\cite{Chisnell}. 

Nontrivial topological phases for Lu$_2$V$_2$O$_7$ are explained theoretically 
by considering a minimal model of spin-1/2 FM Heisenberg model on a 
kagom\'e lattice in the presence of antisymmetric Dzyaloshinskii-Moriya 
(DM) interaction \cite{Dzyaloshinskii,Moriya}, where 
this term acts as a vector potential. In this case, the 
resulting three magnon bands are characterized by $C$=1,0,$-1$, 
where the Zeeman term associated with the magnetic field breaks 
TRS \cite{Li,Sen}. 
The system further exhibits 
topological phase transition (TPT) at the point having zero DM strength. 
Another TMI phase with $C$=$\pm 1$ is obtained later in magnon dispersion of 
two-band honeycomb ferromagnet with DM interaction \cite{Tserkovnya,Owerre}.
For the antiferromagnetic (AFM) case, Heisenberg Shastry-Sutherland model 
with DM interaction exhibits two distinct TMI phases with specific values of 
$C$=$-2,0,2$ and  $C$=$-1,0,1$, obtained for three bosonic triplon dispersion bands   
under longitudinal and cross magnetic fields, respectively \cite{Ganesh,Schmidt}. 

Previous studies reveal that 
values of $C$s for all of the FM bosonic systems are not more than one. 
However, emergence of topological phases with higher 
values of $C$ in fermionic systems 
is found before either invoking distant-neighbor interactions \cite{Piechon} 
or irradiating with circularly polarized light \cite{Arghya}. 
Therefore, search of other bosonic models those are capable to exhibit  
new topological phases as well as phases 
with higher values of $C$ continues in those directions. Recently, 
both AFM and FM models consisting of nearest-neighbor (NN) 
Kitaev \cite{Kitaev}, Heisenberg and symmetric 
spin-anisotropic interactions (KHSA) are introduced those 
are found to exhibit   
TMI phases on the honeycomb lattice in the presence of
magnetic field \cite{Penc,Joshi}. The FM KHSA system hosts 
the TMI phase with $C$=$\pm 1$ \cite{Joshi}. 
Again, value of $C$ is not more than one which is so far true for 
all bosonic lattice models. 

In this study, we report the emergence of a variety TMI  
phases with higher $C$ in FM KHSA model on honeycomb lattice in the 
presence of  next-next-nearest-neighbor (NNNN) interactions in addition to the 
NN terms. Values of $C$s for those 
additional phases in this two-band model are 
$\pm 2$, $\mp 2$, $\pm3$, and $\mp 4$. Higher values of 
$C$ are hereby noticed for the first time on this 
two-band bosonic lattice model. 
However, additional next-nearest-neighbor (NNN) KHSA 
interactions fail to yield 
such new topological phases. 

Interestingly, existence of both NN and NNNN KHSA 
interactions is predicted before in $\alpha$-RuCl$_3$ by combined 
{\em ab initio} and strong-coupling approaches, where spin-1/2 Ru$^{3+}$ ions 
constitute the honeycomb structure \cite{Kee1}. 
Both density functional theory and nonperturbative exact 
diagonalization methods also support the presence of 
NNNN interactions in $\alpha$-RuCl$_3$ \cite{Gong,Valenti}. 
Values of all the interaction strengths have been determined. 
Besides the Kitaev spin liquid ground state, $\alpha$-RuCl$_3$ 
is recently found to demonstrate the half-quantized 
thermal Hall effect \cite{Kasahara}.  
Meanwhile, emergence of multiple TMI phases with 
higher values of $C$ for the model 
based on this compound turns the system more attractive. 

However, the number of 
magnetic compounds those properties are explained in terms of various 
KHSA models is not less. 
KHSA Hamiltonian has been suggested for the 
Kitaev spin-liquid compounds, X$_2$IrO$_3$, X = (Na, Li), 
based on the results of 
both {\em ab initio} and perturbative calculations,  
where the spin-1/2 Ir$^{4+}$ ions form honeycomb structure \cite{Brink,Kee1}. 
Recently, results of angle-dependent FM resonance (FMR) experiments reveal the 
presence of FM KHSA interactions 
between NN spin-3/2 Cr$^{3+}$ ions within 
all chromium halides, CrY$_3$, Y = (Cl, Br, I), 
where Cr$^{3+}$ ions are arranged in a honeycomb lattice \cite{Hammel}. 

The linear spin-wave theory (LSWT) 
of the FM KHSA model on the honeycomb lattice 
is formulated in the subsequent sections. 
LSWT is valid for spin-1/2 but more 
accurate for higher values of spin moment. 
Therefore, TMI phases for all the above compounds can be 
characterized with specific values of $C$ along with transition between them. 
Evidence of chiral edge states supports the values of $C$ of those TMI phases. 
Value of the thermal Hall conductivity (THC) is obtained to study the 
transition between those TMI phases. Article is arranged in the following manner. 
Section \ref{KHSAmodel} describes the model along with the 
determination of classical ground state. 
LSWT is formulated in 
Sec \ref{LSWT}, where analytic expressions of eigenvalues are given. 
Topological phases have been characterized in Sec \ref{CETHC} while 
Sec \ref{discussion} contains discussion on the results.
\section{Kitaev-Heisenberg magnet with spin-anisotropic interactions on honeycomb lattice}
\label{KHSAmodel}
The model comprises Kitaev, Heisenberg and 
spin-anisotropic interactions on the NN  
as well as NNNN bonds in the presence of an 
external magnetic field. The latter two interactions depend on the 
choice of three different links. 
The Hamiltonian is written as
 \begin{equation}
 \begin{aligned}
{\mathcal H}&=J\sum\limits_{\langle ij \rangle} \boldsymbol{S}_i\cdot\boldsymbol{S}_j
\!+\!2\sum\limits_{\langle ij \rangle_\gamma}\!K_\gamma S^\gamma_i S^\gamma_j
\! +\!\sum\limits_{\langle ij \rangle_\gamma}\!\Gamma_\gamma(S^\alpha_i S^\beta_j\!+\!S^\beta_i S^\alpha_j)\\
&+J^\prime\!\!\!\sum\limits_{\langle\langle\langle ij \rangle\rangle\rangle} \!\!\!\boldsymbol{S}_i\cdot\boldsymbol{S}_j 
 \!+2\!\!\!\!\!\sum\limits_{\langle\langle\langle ij \rangle\rangle\rangle_\gamma}\!\!\!\!\!K^\prime_\gamma S^\gamma_i S^\gamma_j
 \!+\!\!\!\!\!\sum\limits_{\langle\langle\langle ij \rangle\rangle\rangle_\gamma}\!\!\!\!\!\Gamma^\prime_\gamma(S^\alpha_i S^\beta_j\!+\!S^\beta_i S^\alpha_j)\\
&- \boldsymbol{h}\cdot\sum\limits_{i}  \boldsymbol{S}_i.
 \label{ham}
\end{aligned}
  \end{equation}
Here, $J$ ($J^\prime$), $K_\gamma$ ($K^\prime_\gamma$) and 
$\Gamma_\gamma$ ($\Gamma^\prime_\gamma$) are the Heisenberg, Kitaev and spin-anisotropic interaction 
strengths respectively for the NN (NNNN) bonds. $\gamma=x,y,z$, which denotes 
three different links of the honeycomb lattice. ${\boldsymbol h}
=g\mu_{\rm B}{\boldsymbol H}$, 
where ${\boldsymbol H}$ is the strength of magnetic field. 
$S^\alpha_i$ is the $\alpha$-th component of spin operator, 
${\boldsymbol S}_i$, at the $i$-th site, where $\alpha=x,y,z$.  
Summations over NN and NNNN  
bonds are denoted by the indices  ${\langle \cdot \rangle}$ and 
${\langle\langle\langle \cdot \rangle\rangle\rangle} $, respectively. 
Periodic boundary condition (PBC) is assumed along both $x$ and $y$ 
directions. 
Both $J$s and $K$s may assume positive (AFM) and negative (FM) values 
separately. 
\begin{figure}[t]
\psfrag{1}{\tiny (a)}
\psfrag{2}{\tiny (b)}
\psfrag{3}{\tiny (c)}
\psfrag{A}{\tiny $A$}
\psfrag{B}{\tiny $B$}
\psfrag{a1}{\tiny $\delta_1$}
\psfrag{a2}{\tiny $\delta_2$}
\psfrag{aa}{\tiny $a$}
\psfrag{x}{\tiny $\hat{x}$}
\psfrag{y}{\tiny $\hat{y}$}
\psfrag{a}{\tiny $K_x$}
\psfrag{b}{\tiny $K_y$}
\psfrag{c}{\tiny $K_z$}
\psfrag{d}{\tiny $K^\prime_x$}
\psfrag{e}{\tiny $K^\prime_y$}
\psfrag{f}{\tiny $K^\prime_z$}
\psfrag{N}{\tiny $N$ units}
\psfrag{k1}{\tiny $k_x$}
\psfrag{k2}{\tiny $k_y$}
\psfrag{G}{\tiny  $\Gamma$}
\psfrag{K}{\tiny K}
\psfrag{M}{\tiny M}
    \begin{minipage}{0.15\textwidth}
     \centering
 \includegraphics[width=189pt]{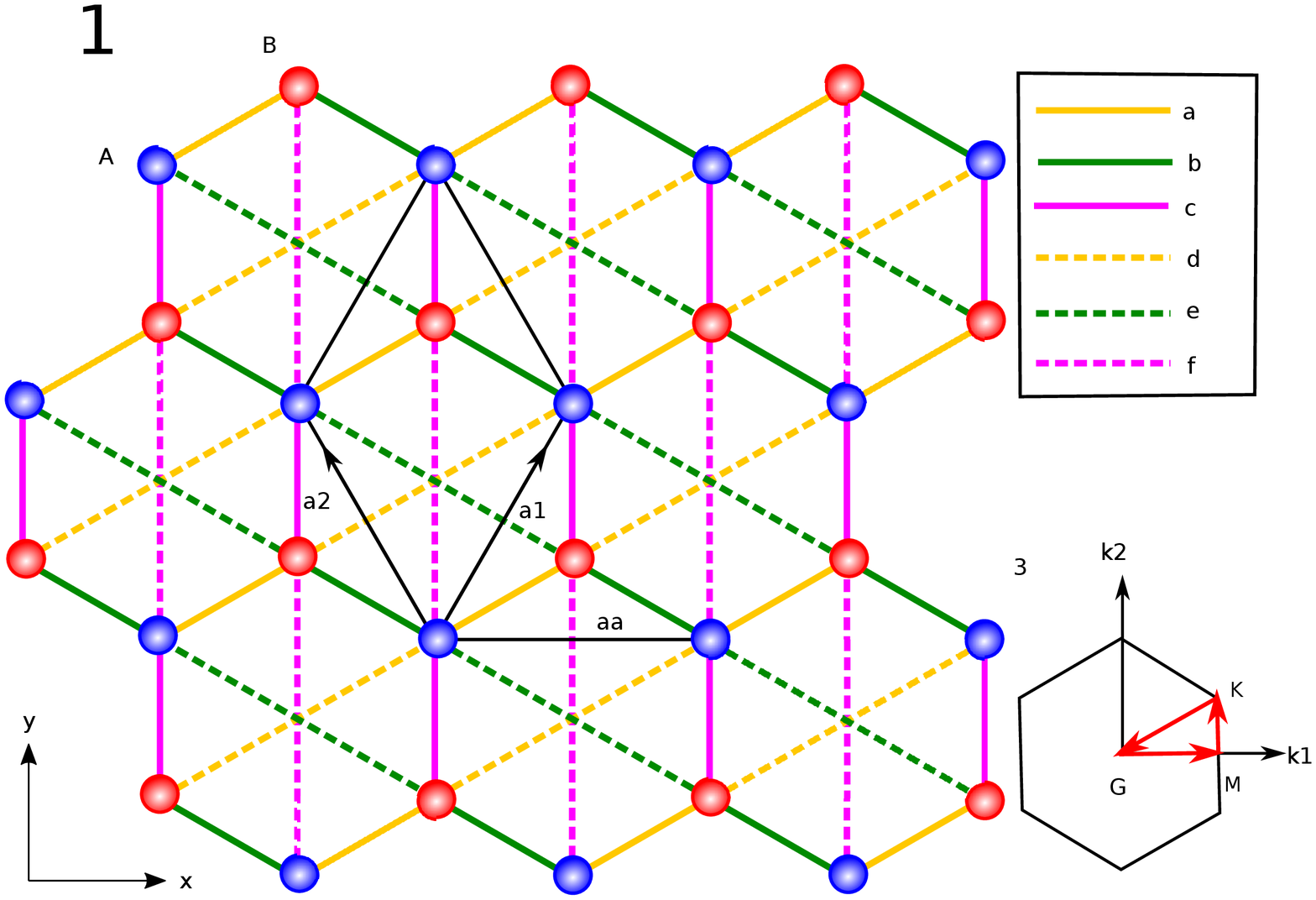}
   \end{minipage}\hfill
   \begin{minipage}{0.10\textwidth}
   \centering
   \includegraphics[width=52pt]{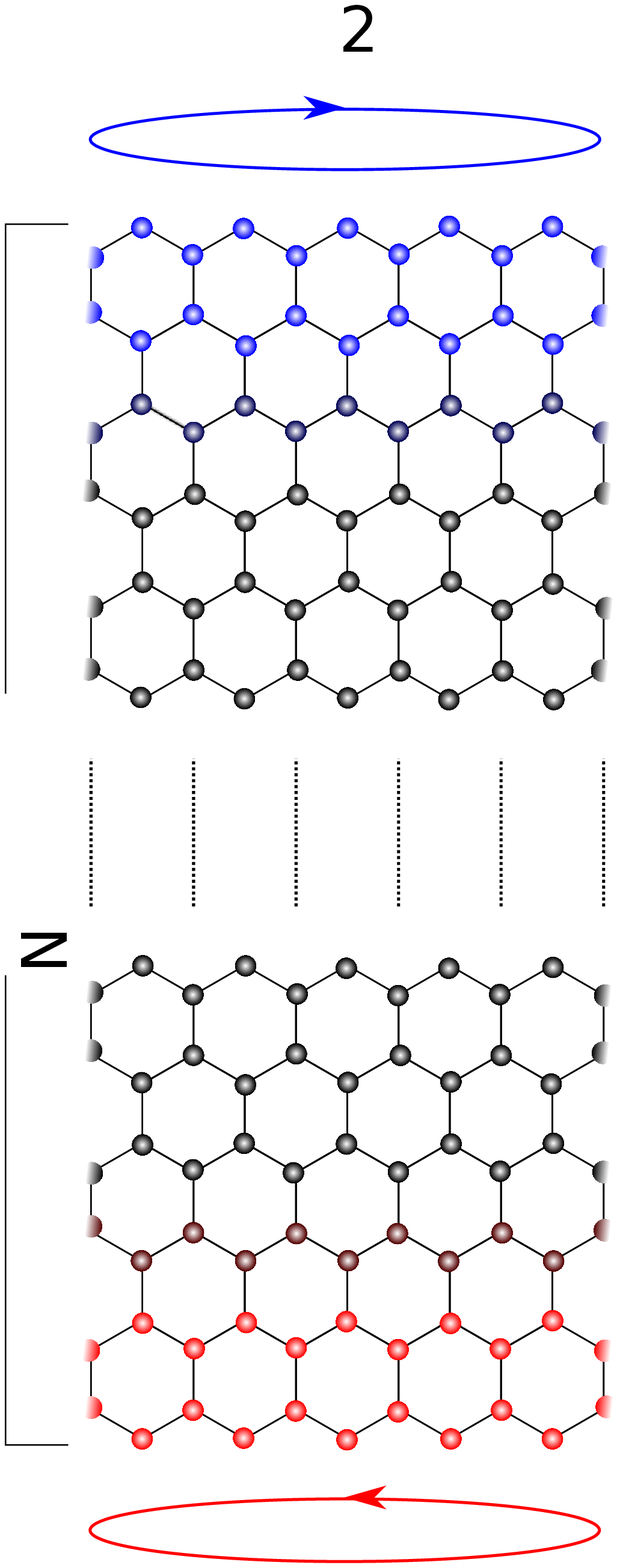}
 \end{minipage}\hfill
  \caption{(color online) (a) Geometrical view of the model with NN and NNNN interactions 
on the honeycomb lattice, (b) geometry of the lattice used for edge 
state calculation, upper and lower edges are drawn in blue and red colors, 
respectively, (c) the first Brillouin zone showing the high-symmetry 
points, $\Gamma$, M and K.}
   \label{lattice}
     \end{figure}  

Geometrical view of this spin model is given in Fig \ref{lattice}(a). 
NN (NNNN) bonds are shown by solid (dashed) lines. 
${\bf \delta_1}$ and ${\bf \delta_2}$ are the primitive vectors of  
non-Bravais honeycomb lattice such that the parallelogram bounded by them 
encloses two nonequivalent sites $A$ and $B$, those are shown 
by blue and red spheres, respectively. 

Zero temperature phase diagram of this model in the absence of  
NNNN interactions ($J^\prime=0$, $K^\prime=0$, $\Gamma^\prime=0$) 
is already studied. For $\Gamma=0$, system exhibits 
four spin-ordered phases, N\'eel, zigzag, FM, stripy and two intermediate 
spin-liquid phases which lie between N\'eel-zigzag and FM-stripy phases.   
FM phase survives for $0.85\pi<\phi<1.5\pi$, $\phi=\arctan{(K/J)}$, 
when $h=0$ \cite{Khaliullin}.  
With the increase of $h$, FM phase is replaced by the 
spin-polarized (SP) phase which begins to spread and 
ultimately occupy the whole parameter space, $0<\phi\leq 2\pi$ 
above the critical value $h_c(\phi)$ \cite{Trebst,Vojta}. 
Value of $h_c$ also depends on the direction of ${\boldsymbol h}$.  
%Zigzag and stripy phases are occupied by 
%the SP phase quickly with the increase of 
%$h$ as they reside around the FM phase. 
SP phase quickly consumes the entire zigzag and stripy phases 
with the increase of 
$h$ as they reside around the FM phase. 
Quantum states for FM and SP phases are 
even though identical but their causes are different. 
However, in this region, the system is topologically trivial. 
 For $\Gamma \neq 0$, additional two spin-ordered phases, 
120$^{\circ}$ order and incommensurate spiral are found to appear in the 
phase diagram when $S=1/2$ and $h=0$ \cite{Kee2}. 
Interestingly, SP phase is topologically nontrivial since the 
magnon dispersion bands emerge with  
$C=\pm 1$ and $\mp 1$, when $h>\Gamma_z$ and directed toward $\hat z$ and [111] 
directions, respectively \cite{Joshi}. 
However, no additional topological phase is found to appear in the 
presence of NNN interactions. 

This work reveals that the system exhibits  
additional four topological phases 
($C=\pm 2, \mp 2, \pm 3, \mp 4$), with the inclusion of 
NNNN terms, when $h>|\Gamma_z+\Gamma_z^\prime|$. 
Therefore, the SP phase hosts six different topological phases 
in this case when $h$ points toward $\hat z$. 
Zigzag order is observed in the compounds 
$\alpha$-RuCl$_3$ and X$_2$IrO$_3$ below the respective 
N\'eel temperatures \cite{Valenti} while FM order is found in 
CrI$_3$ below the Curie temperature \cite{Hammel}. 
So, the results of LSWT on the KHSA model
based on the FM ground state are applicable on those materials 
as long as ${\boldsymbol h}$ is nonzero. 

The extend of FM ground state of this model for the honeycomb lattice 
can be obtained in terms of that for a single unit cell. For this purpose, 
a hexagon is 
considered that holds classical spin vector each at its six vertices.  
Components of the spin vectors have been expressed in terms of the polar coordinates $\theta$ and $\phi$ as
$\boldsymbol{S}_n= S\left(\sin{\theta_n} \cos{\phi_n},\,\sin{\theta_n} \sin{\phi_n}, \cos{\theta_n}\right)$, 
 where $n=1,2,3,4,5,6$.
The minimum value of classical energy corresponds to the
 spin configuration given by $\phi_n=0$ and $\theta_n=0$ 
when both $J$s and $K$s are negative. 
In this case, classical energy of a single hexagon is given by
$E_{\text{\hexagon}}=6\left(J+2K\right)S^2+3\left(J^\prime+2K^\prime\right)S^2-6 h S$,
when $K=K_\gamma$, $K^\prime=K^\prime_\gamma$ and 
$\boldsymbol{h}$ is acting along the $\hat{z}$ direction. 
$E_{\text{\hexagon}}$ does not depend on the 
spin-anisotropic interaction due to the symmetry of the spin configuration. 
Therefore, in the classical ground state 
all the spins orient along the $\hat{z}$ direction when 
both $J$s and $K$s are negative which in the other words corresponds to 
the FM ground state. However, in general, SP phase dominates  
as long as $h/S \geq J+2K+(J^\prime+2K^\prime)/2$. 
%Here, $J$s and $K$s are always kept negative to ensure the 
%FM ground state while obtaining the TMI phases since the whole ground state 
%phase diagram of this model is not available. 
Therefore, in this whole parameter regime single magnon 
dispersions can be treated as exact one. 
\section{Magnon dispersion relations}
\label{LSWT}
In order to obtain the FM magnon dispersion relations based on the 
classical ground state, spin 
operators are expressed in terms of bosonic creation ($b^\dagger$) 
and annihilation $b$ operators through the Holstein-Primakoff transformation as
\begin{equation}
 \begin{aligned}
  S_j^z=S-b_j^\dagger b_j,\;\;
  S_j^+\simeq\sqrt{2S}\, b_j,\;\;
  S_j^-\simeq\sqrt{2S}\, b_j^\dagger. 
 \end{aligned}
\end{equation}
Following the LSWT, Hamiltonian (Eq.\ref{ham}) 
has been expressed in the momentum space by performing the 
Fourier transformation of the operators, 
$b_j=\frac{1}{\sqrt N}\,\sum_{\boldsymbol{k}}b_{\boldsymbol{k}}\,e^{i\,\boldsymbol{k}\cdot 
\boldsymbol{R}_j}$. 
\begin{equation}
 \begin{aligned}
{\mathcal H}=\!\frac{3NS^2}{2}\!\big[J\!+\!J'\!+\!2(K\!+\!K')\big]\!-\!NhS+\!\frac{S}{2}\sum\limits_{\bold{k}}\Psi^\dagger_{\bold{k}}
{\mathcal H}_\bold{k}\Psi_{\bold{k}}, 
 \end{aligned}
\label{hhexa}
 \end{equation}
 where $N$ is the total number of sites and 
$\Psi^\dagger_{\bold{k}}=[b^\dagger_{A,\bold{k}},b^\dagger_{B,\bold{k}},b_{A,-\bold{k}},b_{B,-\bold{k}}]$. 
Now, $b^\dagger_A$ and $b^\dagger_B$ are the bosonic creation operators 
on sublattices $A$ and $B$, respectively. 
The estimated values of $\Gamma$ and $\Gamma^\prime$ for the Kitaev materials 
are generally very small. 
For $\alpha$-RuCl$_3$ they are given by $\Gamma/K\approx 0.7$ 
and $\Gamma^\prime/K\approx 0.0$ \cite{Kee1}. $\Gamma/K\approx 0.013$ and 0.0 
for CrI$_3$ \cite{Hammel} and X$_2$IrO$_3$ \cite{Valenti}, respectively.  
On the other hand, topological phases appear as soon as values of 
 $\Gamma$ and $\Gamma^\prime$ are nonzero, and they are found robust against 
any alteration of them. 
Therefore, one can assume both $x$ and $y$ components of 
$\Gamma$ and $\Gamma^\prime$ are negligible for the sake of analytic  
convenience. Otherwise, Eq \ref{hhexa} includes linear and cubic terms 
of $b$ operators. It is worth mentioning that 
different TMI phases appear for different values of $J$s and $K$s 
and a definite phase survives as long as $h>|\Gamma+\Gamma^\prime|$ 
without depending 
on the individual values of $h$, $\Gamma$ and $\Gamma^\prime$. 
Thus by keeping only the bilinear terms, form of the $4\times 4$ matrix, 
${\mathcal H}_\bold{k}$ is given below.
 \begin{equation}
 \begin{aligned}
{\mathcal H}_\bold{k}=
 \left[ 
 { \begin{array}{cc}
  X_\bold{k}  &  Y_\bold{k}  \\
  Y^\dagger_\bold{k} &   X^T_{-\bold{k}}  \\
  \end{array}}
\right]\!,\,
X_\bold{k}=
 \left[ 
 { \begin{array}{cc}
  a_0  &   a_{\bold{k}}  \\
  a^*_{\bold{k}} &   a_0  \\
  \end{array}}
\right]\!, \,
Y_\bold{k}=
 \left[ 
 { \begin{array}{cc}
  0  &   b_{\bold{k}}  \\
  b_{-\bold{k}} &   0  \\
  \end{array}}
\right]\!,
\label{Hk}
 \end{aligned}
 \end{equation}
where, 
\begin{equation}
  \begin{aligned}
  a_0=&-3J-2K-3J^\prime-2K^\prime+h/S, \\
  a_{\bold{k}}=&J+(J+K)\left(e^{-i\bold{k}\cdot\boldsymbol{\delta}_1}+e^{-i\bold{k}\cdot\boldsymbol{\delta}_2}\right)\\
&+J^\prime e^{-i\bold{k}\cdot\boldsymbol{n}_1}+(J^\prime+K^\prime)\left(e^{-i\bold{k}\cdot\boldsymbol{n}_2}+e^{-i\bold{k}\cdot\boldsymbol{n}_3}\right), \\
  b_{\bold{k}}=&K\left(e^{-i\bold{k}\cdot\boldsymbol{\delta}_1}-e^{-i\bold{k}\cdot\boldsymbol{\delta}_2}\right)+K^\prime\left(e^{-i\bold{k}\cdot\boldsymbol{n}_3}-e^{-i\bold{k}\cdot\boldsymbol{n}_2}\right)\\
&+i\Gamma_z+i\Gamma^\prime_z e^{-i\bold{k}\cdot\boldsymbol{n}_1},\;{\rm with}, \\
\boldsymbol{\delta}_1=&a\left( \frac{1}{2}\,\hat{i}+\frac{\sqrt{3}}{2}\,\hat{j}\right), \;
  \boldsymbol{\delta}_2=a\left(-\frac{1}{2}\,\hat{i}+\frac{\sqrt{3}}{2}\,\hat{j}\right), \\
 \boldsymbol{n}_1=&\sqrt{3}\,a\,\hat{j},\;
 \boldsymbol{n}_2=a\,\hat{i},\;
 \boldsymbol{n}_3=-a\,\hat{i}.\nonumber
\nonumber
  \end{aligned}
\end{equation} 
Here, $a$ is the lattice distance between two adjacent 
$A$ or $B$ sublattice sites 
which is henceforth set to 1. 

\begin{figure}[t]
%  \psfrag{AA}{(a)}
 \psfrag{E}{\tiny $E$}
 \psfrag{kx}{\tiny $k_x$}
 \psfrag{ky}{\tiny $k_y$}
%     \begin{minipage}{0.16\textwidth}
%     \psfrag{c1}{\tiny $C_1=-1$}
% \psfrag{c2}{\tiny $C_2=+1$}
% \centering
%   \includegraphics[width=120pt]{energy_3d_(mp1).eps}
%   \end{minipage}\hfill
   \begin{minipage}{0.18\textwidth}
   \psfrag{BB}{(a)}
   \psfrag{c1}{\tiny $C_1=+1$}
\psfrag{c2}{\tiny $C_2=-1$}
   \centering
  \includegraphics[width=190pt]{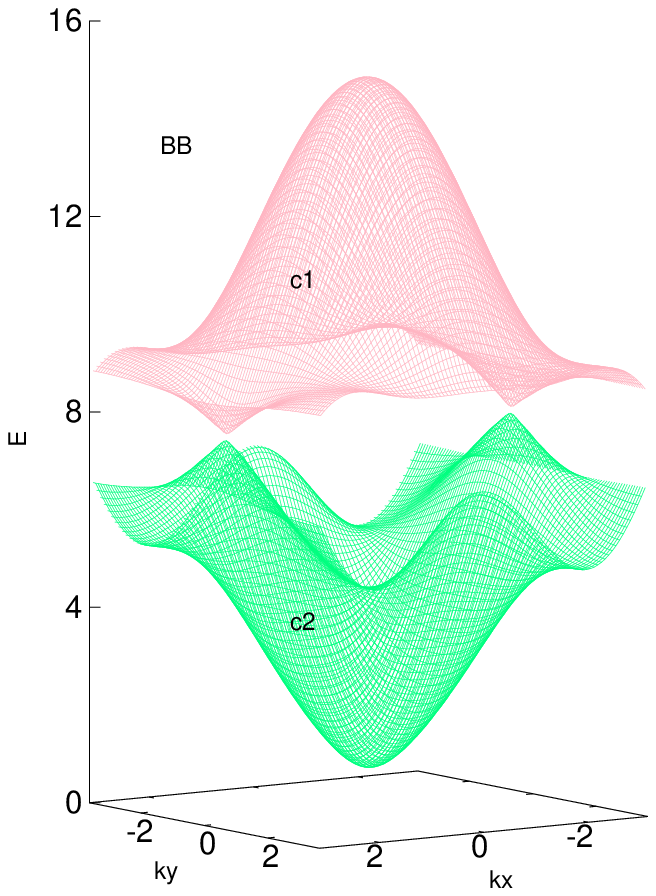}
 \end{minipage}\hfill   
  \begin{minipage}{0.30\textwidth}
  \psfrag{E}{}
  \psfrag{CC}{(b)}
    \psfrag{c1}{\tiny $C_1=-2$}
\psfrag{c2}{\tiny $C_2=+2$}
    \centering
  \includegraphics[width=190pt]{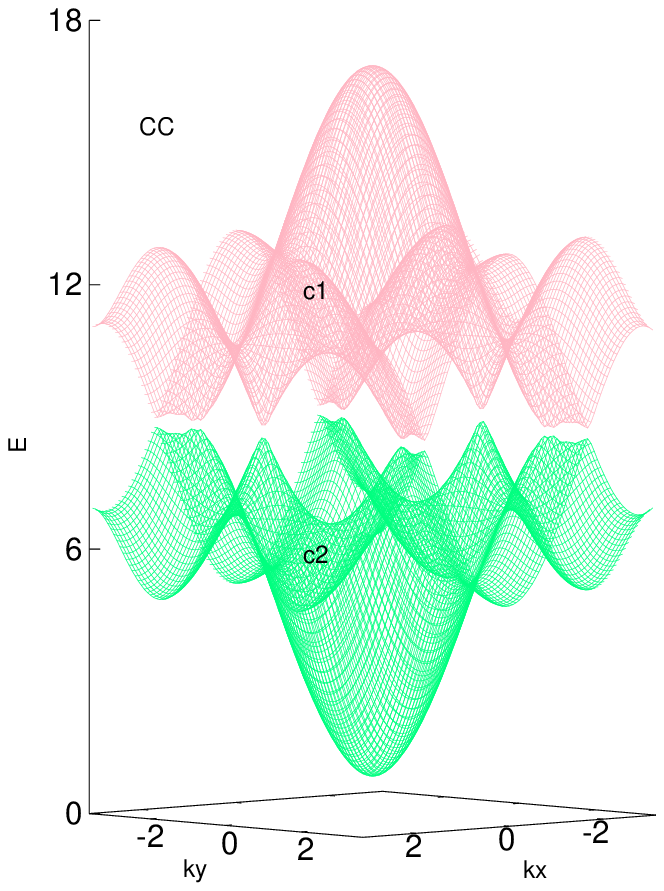}
  \end{minipage}\hfill
  \vskip 0.5cm
%     \begin{minipage}{0.16\textwidth}
%    \psfrag{DD}{(d)}
%    \psfrag{c1}{\tiny $C_1=+2$}
% \psfrag{c2}{\tiny $C_2=-2$}
%    \centering
%   \includegraphics[width=120pt]{energy_3d_(pm2).eps}
%  \end{minipage}\hfill
  \begin{minipage}{0.18\textwidth}
  \psfrag{E}{\tiny $E$}
\psfrag{EE}{(c)}
    \psfrag{c1}{\tiny $C_1=+3$}
\psfrag{c2}{\tiny $C_2=-3$}
    \centering
  \includegraphics[width=190pt]{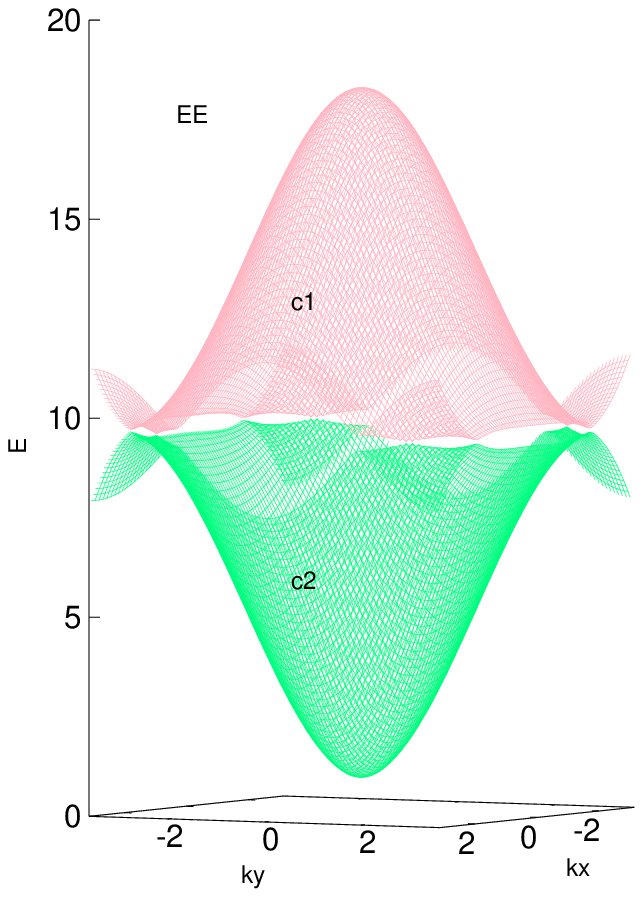}
  \end{minipage}\hfill 
\begin{minipage}{0.30\textwidth}
\psfrag{E}{}
\psfrag{EE}{(d)}
    \psfrag{c1}{\tiny $C_1=-4$}
\psfrag{c2}{\tiny $C_2=+4$}
    \centering
  \includegraphics[width=190pt]{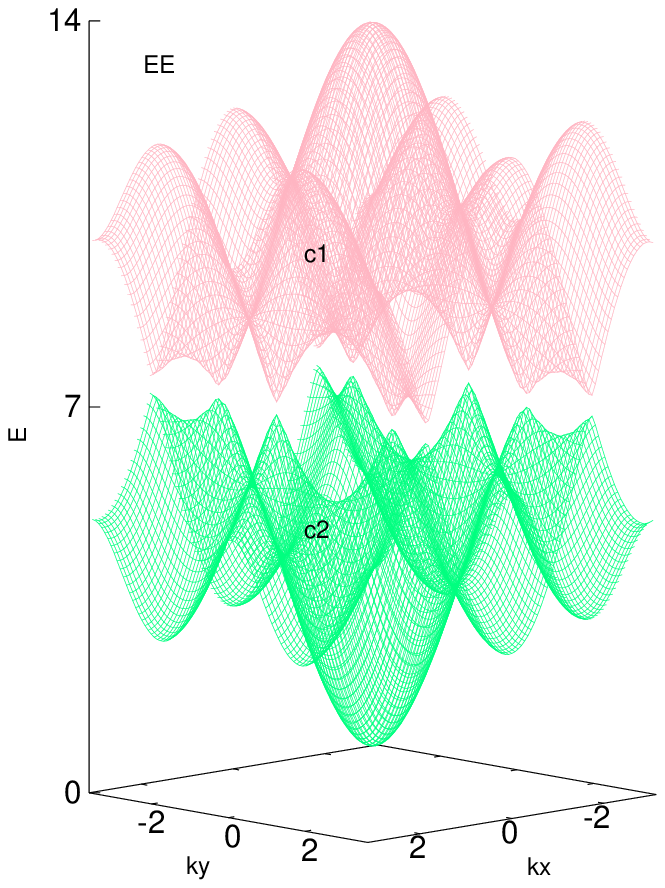}
  \end{minipage}\hfill 
%\vskip 1.1cm
  \caption{(color online) Three-dimensional magnon bands. 
(a) $J=-1$, $K=-1$, $K'=-1$, $\Gamma^\prime_z=0.5$ for $C=\pm 1$, 
(b) $J=-1$, $J^\prime=-1$, $K'=-1$, $\Gamma^\prime_z=0.5$ for $C=\mp 2$, 
(c) $J=-2$, $K=-0.2$, $J^\prime=-1$, $K^\prime=0.4$, $\Gamma^\prime_z=0.5$ for $C=\pm 3$, and 
(d) $J=-0.5$, $J^\prime=-1$, $K^\prime=-1.0$, $\Gamma^\prime_z=0.5$ for $C=\mp 4$. 
No value is specified for those parameters when they are zero.}
   \label{energy3d}
  \end{figure}

According to the bosonic Bogoliubov transformation, the non-Hermitian
matrix $ I_{\rm B} {\mathcal H}_\bold{k}$ is diagonalized in order to 
obtain the eigenenvalues and eigenmodes where $I_{\rm B}=$diag[$1,1,-1,-1$].
The characteristic equation in terms of the eigenvalue, $E_{\bf k}$, is given by 
\(E_{\bf k}^4 + p_{\bf k}\,E_{\bf k}^2+ q_{\bf k}=0\), 
where $p_{\bf k}=|b_{\bf k}|^2+|b_{-\bf k}|^2-2\,(a_0^2+|a_{\bf k}|^2)$  
and $q_{\bf k}=(a_0^2-|a_{\bf k}|^2)^2-a_0^2 \,(|b_{\bf k}|^2+|b_{-\bf k}|^2) 
+|b_{\bf k}|^2\,|b_{-\bf k}|^2-a_{\bf k}^2\, b_{\bf k}^*\,b_{-\bf k}
- a_{\bf k}^{*2} \,b_{\bf k}\, b_{-\bf k}^*$. 
Real eigenvalues are found as long as $h>|\Gamma_z+\Gamma_z^\prime|$. 
Magnon dispersion relations 
are given by the positive eigenvalues of the characteristic equation,  
$E_{\bf k}^\pm=[(-p_{\bf k}\pm \sqrt{p_{\bf k}^2-4q_{\bf k}})/2]^\frac{1}{2}$. 
$E_{\bf k}^\pm$ with specific values of $C$ have been plotted 
in Fig \ref{energy3d} for four different TMI phases, 
(a) $C=\pm 1$, (b) $C=\mp 2$, (c) $C=\pm 3$, and (d) $C=\mp 4$. 

   \begin{figure}[t]
  \psfrag{E}{\tiny $E$}
\psfrag{K}{\tiny K}
\psfrag{G}{\tiny  $\Gamma$}
\psfrag{K1}{\footnotesize $\bold{k}$}
\psfrag{M}{\tiny M}
%\psfrag{DOS}{\tiny DOS}
% 
%   %\begin{minipage}{0.16\textwidth}
%   \psfrag{AA}{\scriptsize(a)}
%   \psfrag{C1}{\tiny $C_1=-1$}
% \psfrag{C2}{\tiny $C_2=+1$}
%   \centering
%   \includegraphics[width=90pt]{joshi_nnnn_bz_(mp1).eps}
%   \end{minipage}\hfill
    \begin{minipage}{0.21\textwidth}
   \psfrag{BB}{\scriptsize(a)}
   \psfrag{C1}{\tiny $C_1=+1$}
\psfrag{C2}{\tiny $C_2=-1$}
    \centering
  \includegraphics[width=128pt]{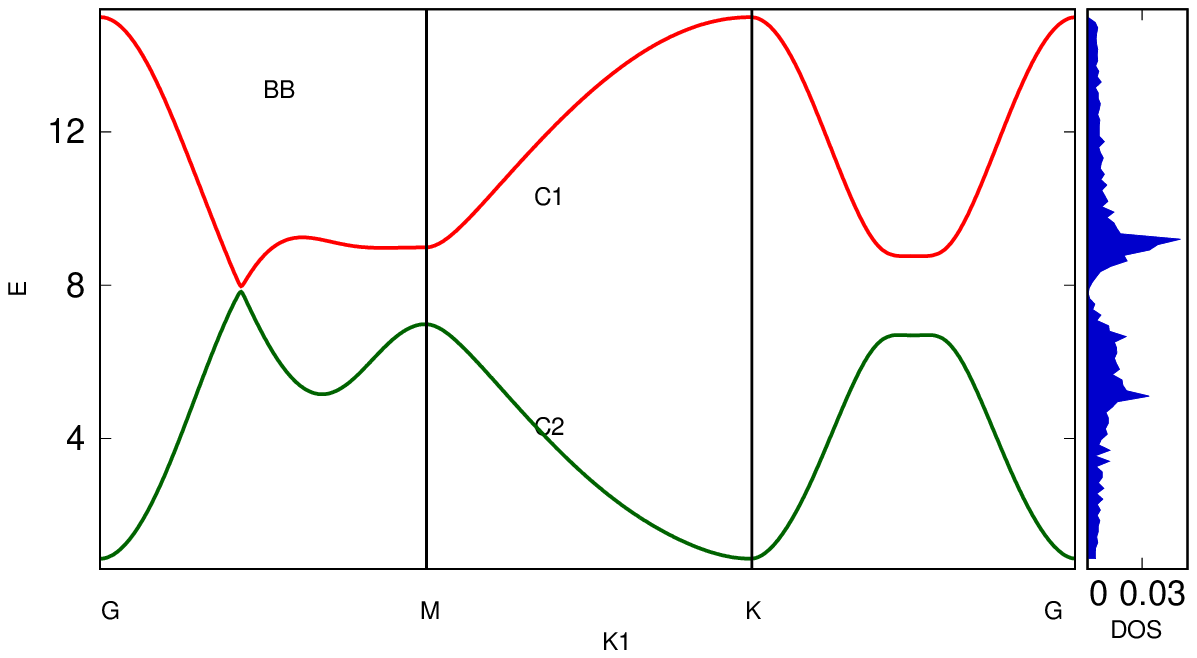}
  \end{minipage}\hfill   
  \begin{minipage}{0.25\textwidth}
   \psfrag{CC}{\scriptsize(b)}
    \psfrag{C1}{\tiny $C_1=-2$}
\psfrag{C2}{\tiny $C_2=+2$}
     \centering
  \includegraphics[width=128pt]{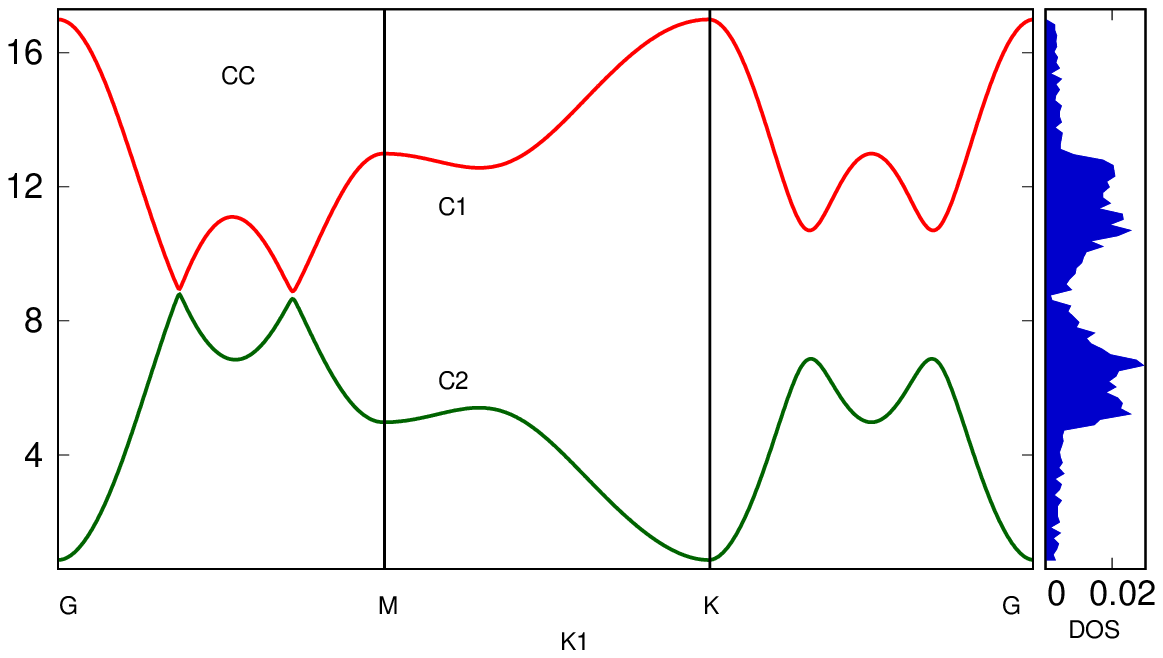}
   \end{minipage}\hfill    
  
  \vskip -0.70cm
  
     \begin{minipage}{0.21\textwidth}
   \psfrag{EE}{\scriptsize(c)}
  \psfrag{C1}{\tiny $C_1=+3$}
\psfrag{C2}{\tiny $C_2=-3$}
    \centering
  \includegraphics[width=128pt]{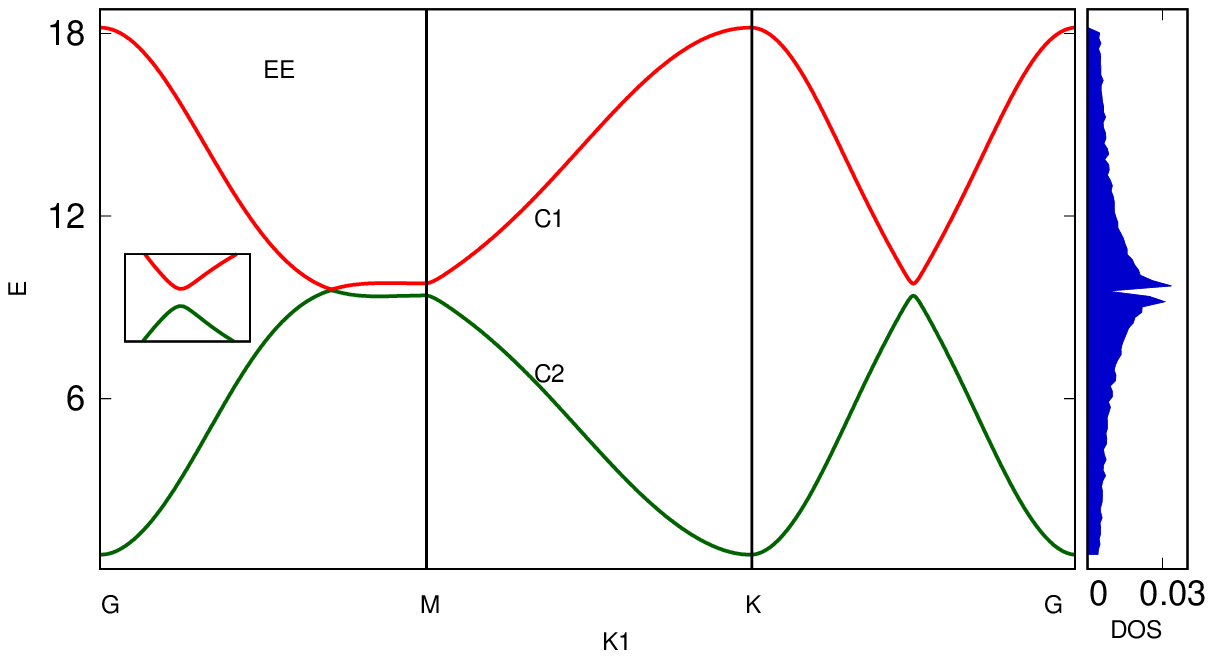}
  \end{minipage}\hfill   
  \begin{minipage}{0.25\textwidth}
 \psfrag{EE}{\scriptsize(d)}
  \psfrag{C1}{\tiny $C_1=-4$}
\psfrag{C2}{\tiny $C_2=+4$}
     \centering
  \includegraphics[width=128pt]{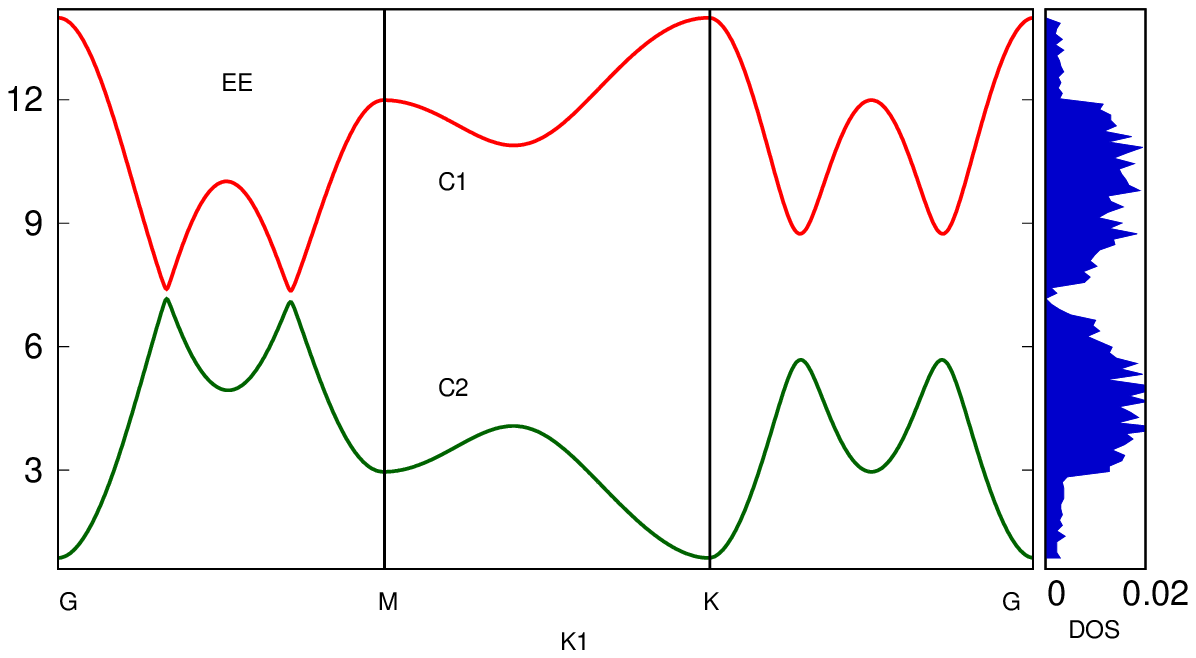}
   \end{minipage}\hfill   
   \vskip -0.70cm
  \caption{(color online) Dispersion relation along the high-symmetry points of Brillouin zone. 
The side panel shows the density of states. Values of the parameters are the 
same as Fig \ref{energy3d} for the respective plots.}
     \label{energybz}
  \end{figure}  
Number of band touching points is the minimum in 
 Fig \ref{energy3d} (a) where $C=\pm 1$, while it is the 
maximum in (d) when $C=\mp 4$. Dispersion relations along the 
high-symmetry points of BZ in addition to the density of states (DOS) are 
shown in Fig \ref{energybz} for four different cases. DOS 
reveals the presence of true band gap in every case. 
The presence of NNN interactions in this specific model 
only alter the diagonal  
elements of the Eq \ref{Hk}. Thus it fails to generate additional TMI phase since 
the values of $C$ are found technically insensitive to 
the values of diagonal components.   
\section{Chern numbers, edge states and thermal hall conductance}
\label{CETHC}
In order to draw the topological phase diagram, 
value of Chern number and edge states have been obtained 
throughout the parameter space of the system. 
Chern number of a particular band has been calculated by
integrating the Berry curvature over the Brillouin zone (BZ).
\begin{equation}
  \begin{aligned}
   C=\frac{1}{2\pi }\iint_{BZ} F(\bold{k})dk_x dk_y,
  \end{aligned}
\end{equation}
where the Berry curvature of that band, $F(\bold{k})$, is expressed 
in terms of the corresponding Berry connection, 
$A_{\mu}(\bold{k}) = \braket{n(\bold{k})|\partial_{k_{\mu}}|n(\bold{k})}$ as 
$F (\bold{k})= {\partial_{ k_x}} A_{y}(\bold{k})-
{\partial_{ k_y}} A_{x}(\bold{k})$, and 
$|n(\bold{k})\rangle$ is the eigenvector of that particular magnon band. 
Ultimately, the value of $C$ has been estimated numerically following 
the method proposed by Fukui and others \cite{Fukui}. 

  \begin{figure}[t]
 \psfrag{E}{\tiny $E$}
 \psfrag{qx}{\tiny $k_x$}
 \psfrag{w}{\tiny $|\psi^2|$}
 \psfrag{AA}{\scriptsize(a)}
 \psfrag{BB}{\scriptsize(a)}
\psfrag{CC}{\scriptsize(b)}
 \psfrag{DD}{\scriptsize(c)}
 \psfrag{EE}{\scriptsize(d)}
 \psfrag{site}{\tiny site}
 %\psfrag{DOS}{\tiny DOS}
 \psfrag{ub}{\tiny upper edge}
 \psfrag{lb}{\tiny lower edge}
%     \centering
%      \begin{minipage}{0.23\textwidth}
%  \includegraphics[width=128pt]{edge_state_energy_n_DOES_joshi_two_plots_(mp1).eps}
%    \end{minipage}\hfill
    \begin{minipage}{0.21\textwidth}
    \centering
   \includegraphics[width=128pt]{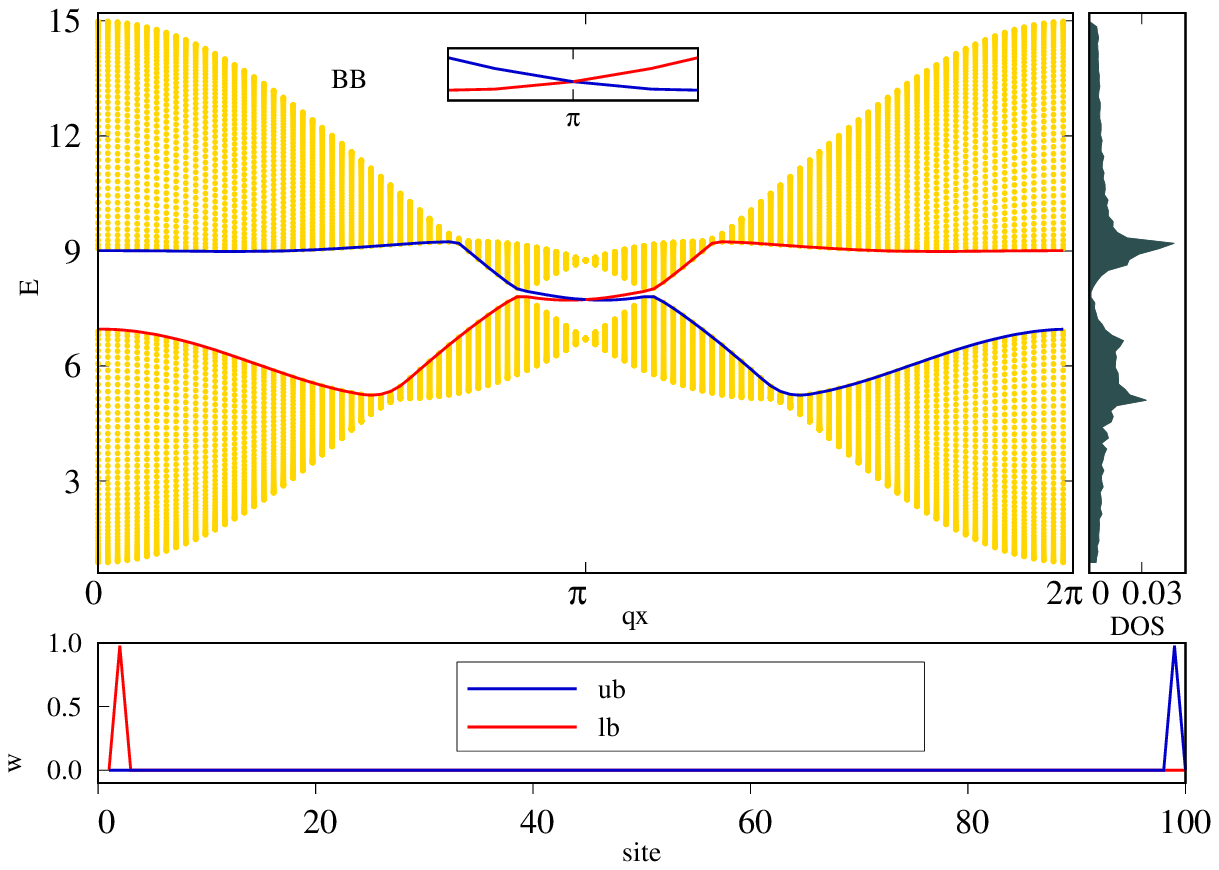}
  \end{minipage}\hfill
   \psfrag{E}{}
   \psfrag{w}{}
    \centering
  \begin{minipage}{0.245\textwidth}
   \includegraphics[width=128pt]{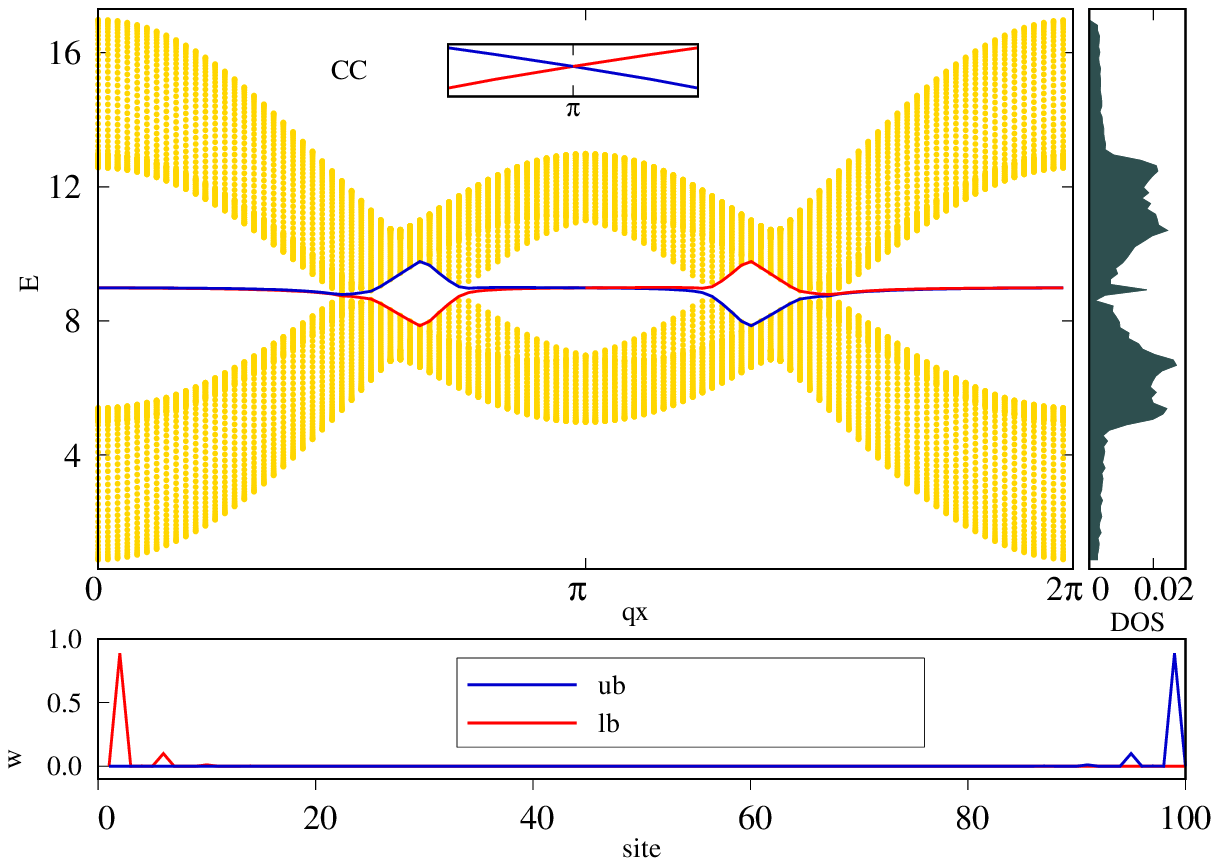}
   \end{minipage}\hfill
     \vskip 0.07cm
    \begin{minipage}{0.21\textwidth}
     \psfrag{E}{\tiny $E$}
   \psfrag{w}{\tiny $|\psi^2|$}
    \centering
   \includegraphics[width=128pt]{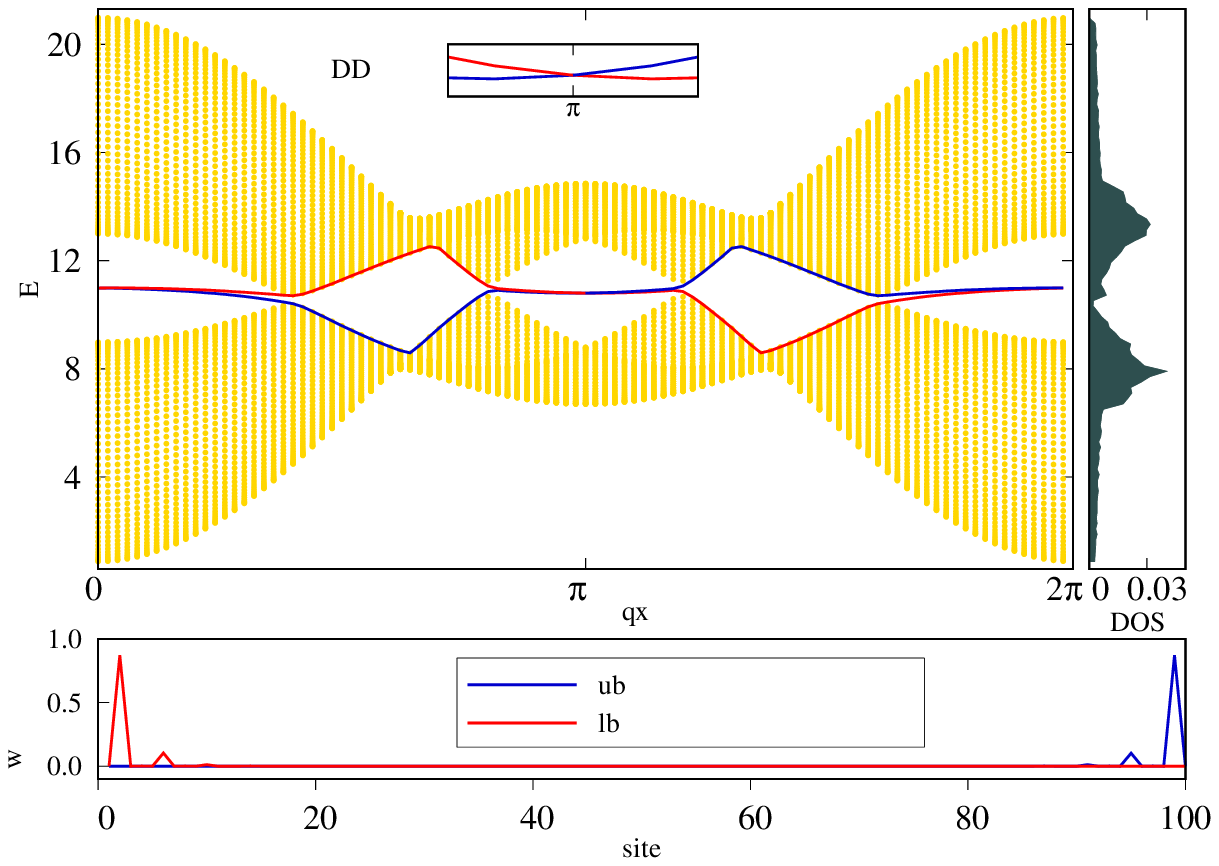}
  \end{minipage}\hfill
   \begin{minipage}{0.245\textwidth}
    \psfrag{E}{}
   \psfrag{w}{}
   \centering
   \includegraphics[width=128pt]{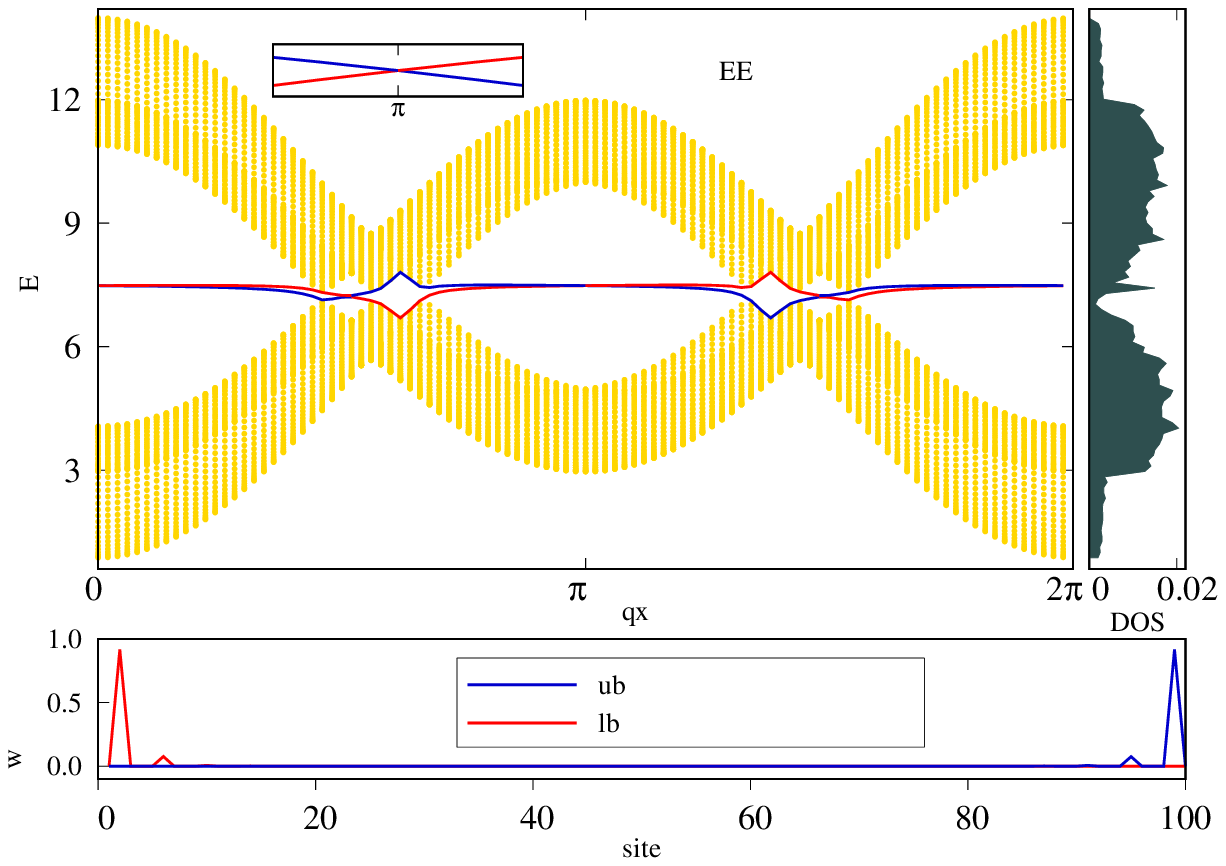}
   \end{minipage}\hfill
  \caption{(color online) Magnon dispersions of edge and bulk states in the one-dimensional BZ. 
Upper and lower edge modes are drawn in blue and red lines, respectively, while 
bulk modes are in golden points. Insets show the closer view of crossing points of edge energies 
around a definite value of $k_x$.
The side panel shows the density of edge states.
 The lower panel indicates variation of probability density of both edge
modes with respect to site number for a fixed $k_x$. Values of the parameters are:  
(a) $J=-1$, $K=-1$, $K'=-1$, $\Gamma^\prime_z=0.5$ for $C=\pm 1$, 
(b) $J=-1$, $J^\prime=-1$, $K'=-1$, $\Gamma^\prime_z=0.5$ for $C=\mp 2$, 
(c) $J=-1$, $K=-1$, $J^\prime=-1$, $K'=-1$, $\Gamma^\prime_z=0.5$ for $C=\pm 2$, and 
(d) $J=-0.5$, $J^\prime=-1$, $K^\prime=-1.0$, $\Gamma^\prime_z=0.5$ for $C=\mp 4$. 
No value is assigned to those parameters when they are zero.}
 \label{edge}
   \end{figure}  
 \begin{figure}[h]
 \psfrag{E}{\tiny $E$}
 \psfrag{qx}{\tiny $k_x$}
 \psfrag{w}{\tiny $|\psi^2|$}
 \psfrag{AA}{\scriptsize(a)}
 \psfrag{BB}{\scriptsize(b)}
 \psfrag{site}{\tiny site}
 %\psfrag{DOS}{\tiny DOS}
 \psfrag{ub}{\tiny upper edge}
 \psfrag{lb}{\tiny lower edge}
  \begin{minipage}{0.21\textwidth}
   \centering
   \includegraphics[width=128pt]{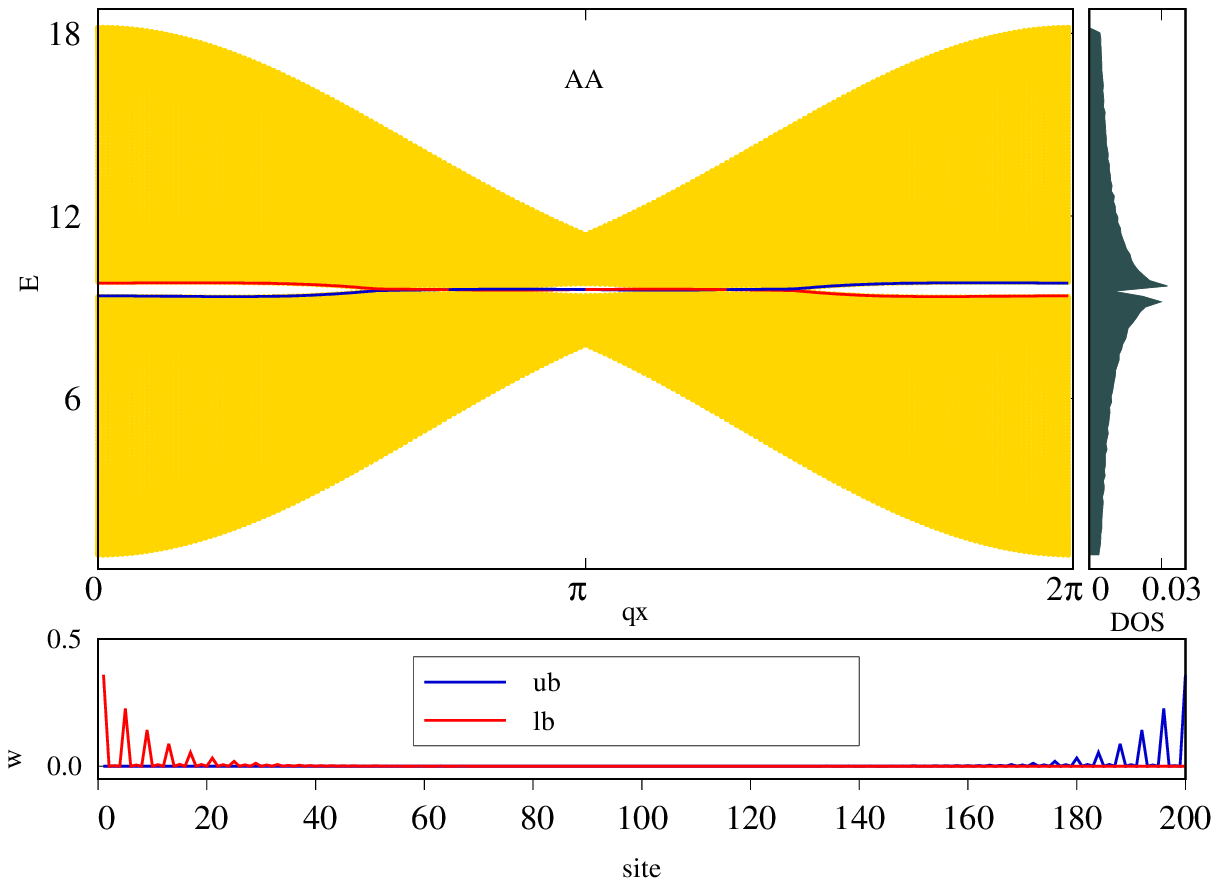}
  \end{minipage}\hfill
   %\hspace{0.01cm}
   \begin{minipage}{0.24\textwidth}
   \psfrag{E}{}
  \centering
   \includegraphics[width=128pt]{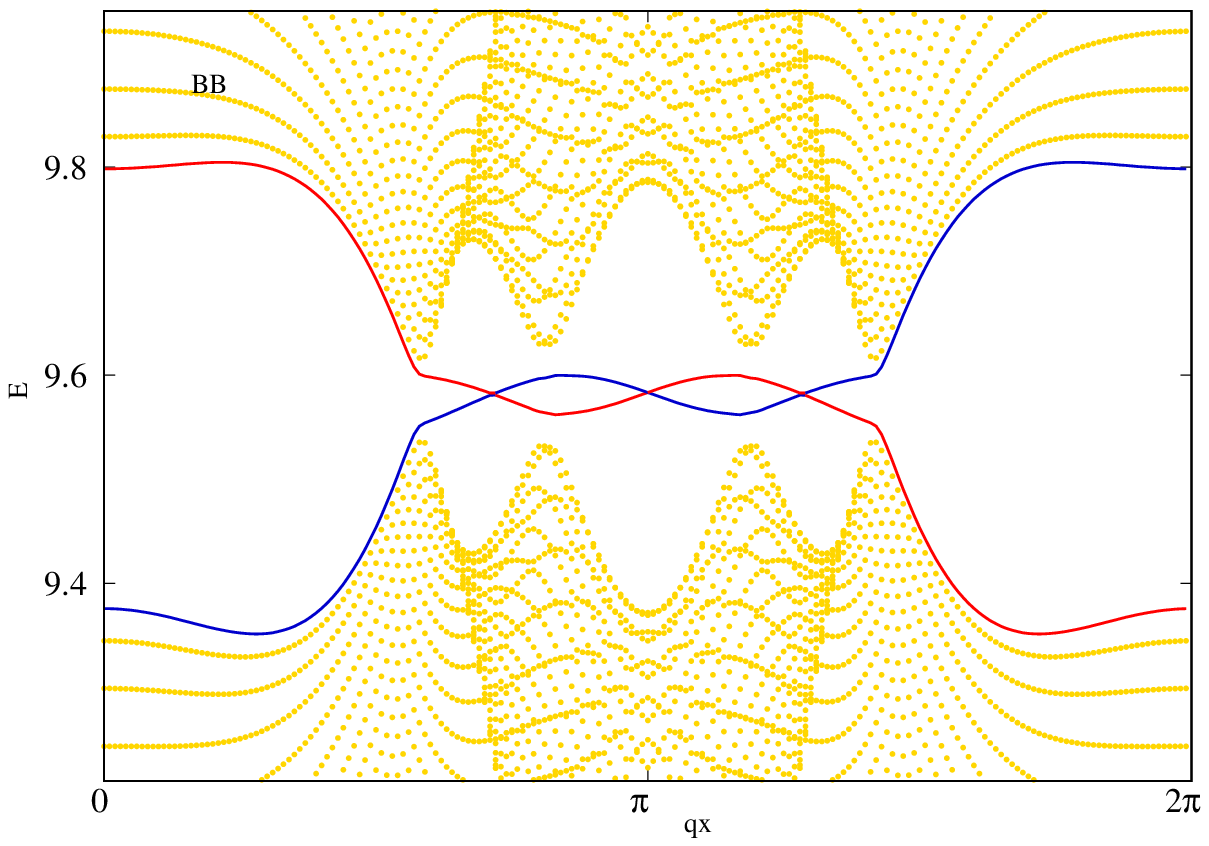}
   \end{minipage}\hfill
  \caption{(color online) (a) Bulk and edge magnon mode energies showing the existence of 
TMI phase with $C=\pm 3$, when $J=-2$, $K=-0.2$,  $\Gamma_z=0.0$, 
$J^\prime=-1$, $K^\prime=0.4$, and $\Gamma^\prime_z=0.5$, 
(b) a closer view of the three crossing points around $9<E<10$. }
\label{edge1}
   \end{figure}  
 \begin{figure*}[t]
    \psfrag{1}{\tiny I}%$C_1$=+4}
\psfrag{2}{\tiny II}%$C_1$=+2}
\psfrag{3}{\tiny III}%$C_1$=+1}
\psfrag{4}{\tiny IV}%$C_1$=0}
\psfrag{5}{\tiny V}%$C_1$=-1}
\psfrag{6}{\tiny VI}%$C_1$=-2}
\psfrag{a}{\tiny $(\mp4)$}
\psfrag{b}{\tiny $(\mp2)$}
\psfrag{c}{\tiny $(\mp1)$}
\psfrag{d}{\tiny $(0)$}
\psfrag{e}{\tiny $(\pm1)$}
\psfrag{f}{\tiny $(\pm2)$}
\psfrag{k1}{\tiny $K$}
\psfrag{j2}{\tiny $J^\prime$}
\psfrag{AA}{\tiny(a)}
\psfrag{BB}{\tiny(b)}
\psfrag{CC}{\tiny(c)}
    \begin{minipage}{0.29\textwidth}
  \includegraphics[width=161.2pt]{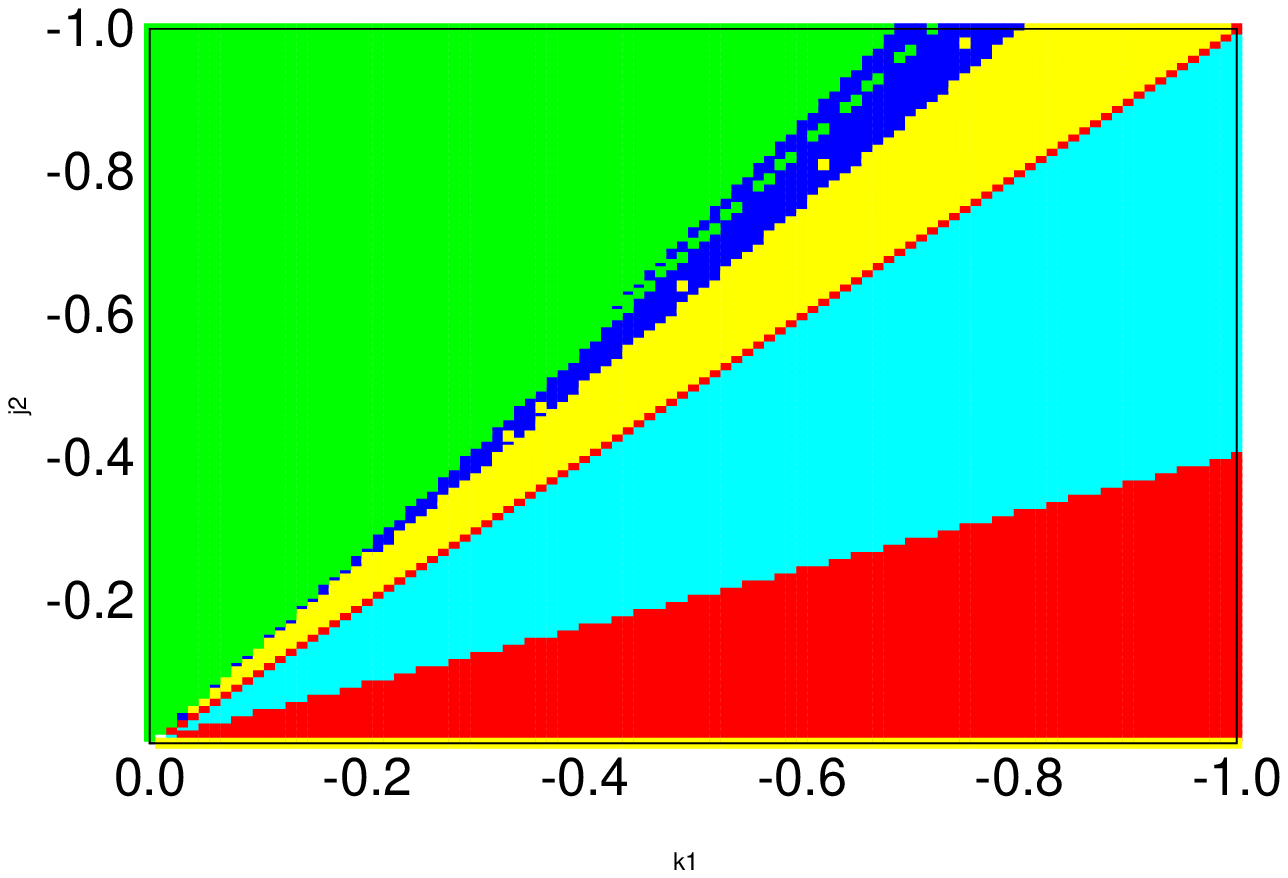}
  \end{minipage}\hfill
  \begin{minipage}{0.29\textwidth}
  \includegraphics[width=161.2pt]{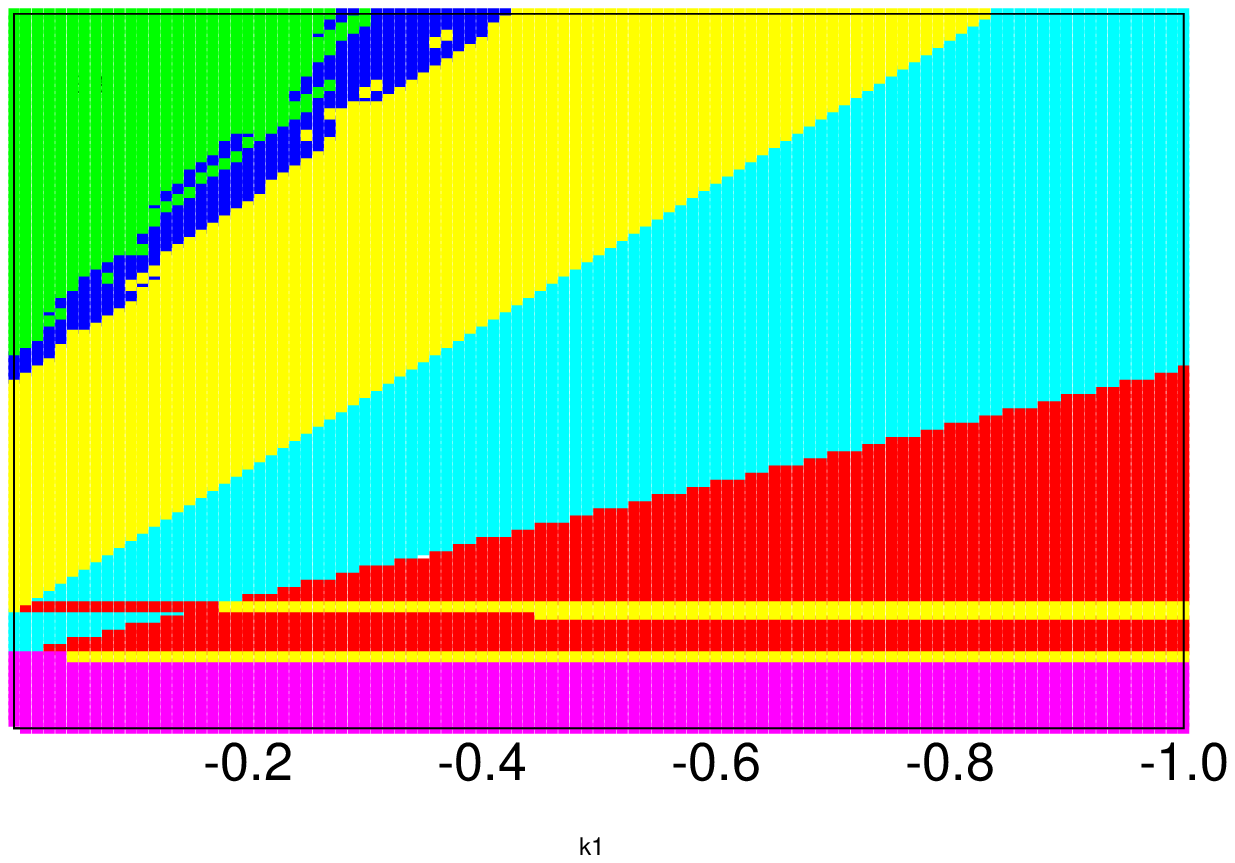}
 \end{minipage}\hfill
   \begin{minipage}{0.29\textwidth} 
   \includegraphics[width=161.2pt]{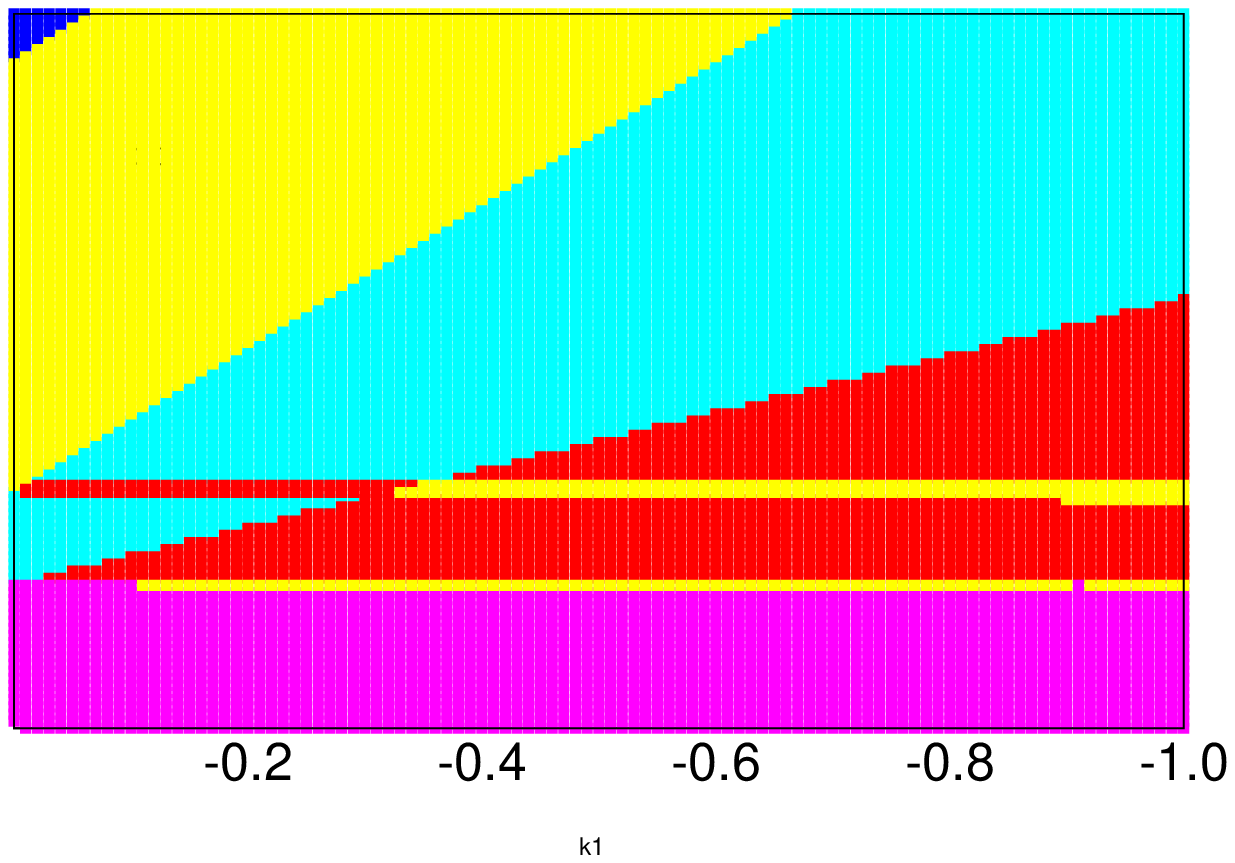}
 \end{minipage}\hfill
  \begin{minipage}{0.11\textwidth}
 \centering
 \includegraphics[width=46.2pt]{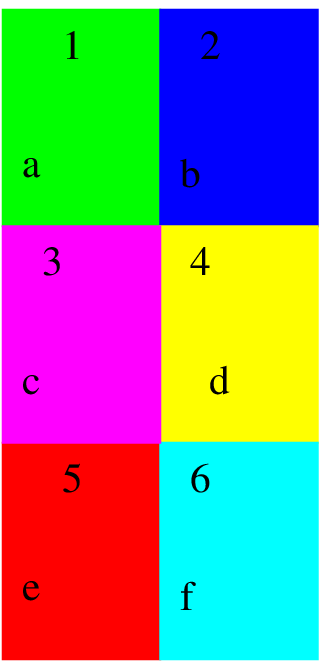}
 \end{minipage}\hfill
  \caption{(color online) Regions of TMI phases of the system in $J'$-$K$ parameter space 
for (a) $J=0$, (b) $J=-0.5$, (c) $J=-1$.}
  \label{topological_phase}
  \end{figure*}  
Nonzero value of $C$ implies the existence of edge states according to the 
BBC rule. Edge states correspond to the surface property of the system. 
A pair of edges parallel to the $x$-axis has been created by breaking the 
PBC along the $y$-axis. For this purpose,  
a strip of honeycomb lattice is considered which has $N$ unit cells 
along the $\hat{y}$ and infinitely long 
towards the $\hat{x}$. This fact is taken into 
account by retaining the PBC along $x$ axis where $k_x$ is 
treated as a good quantum number. Subsequently, taking Fourier
transform of the bosonic operators only along the $x$ direction,  
2$N\times 2N$ Hamiltonian has been obtained.
Diagonalizing the Hamiltonian for $N=50$, eigenvalues and eigenvectors  
of the bulk as well as edge states have been obtained. 

Dispersions of edge and bulk states in the one-dimensional BZ 
for four different topological phases have been shown in Fig \ref{edge}. 
Edge state dispersion branches for upper (blue) and lower (red) sides are 
drawn in different colors. For this two-band system 
number of crossing between those two dispersion branches is actually 
proportional to the absolute value of $C$ for those respective bands. 
TMI phase for $J=-1$, $K=-1$, $\Gamma_z=0$, $J^\prime=0$, $K^\prime=0$, $\Gamma^\prime_z=0.5$ corresponds to $C=\mp 1$, 
 where the upper sign denotes the value of  
$C$ for upper energy band. 
Exchange of Chern numbers occurs when NNNN Kitaev interaction becomes 
equal to $-1$ ($K^\prime=-1$), leading to another TMI phase with $C=\pm 1$, 
which is shown in Fig \ref{edge} (a). 
TMI phase with $C=\pm 2$ appears when $J^\prime=-1$ which is followed by another TMI phase with 
$C=\mp 2$ when NN Kitaev interaction is made zero ($K=0$). 
Those are shown in Figs \ref{edge} (c) and (b), respectively. 
System exhibits TMI phase with $C=\mp 4$ when $J$ is fixed at $-0.5$ 
keeping other 
parameters unchanged, which is shown in Fig \ref{edge} (d). 
Fig \ref{edge1} (a) demonstrates the 
dispersion bands of another TMI phase with $C=\pm 3$,
where the closer view of three distinct crossing points of 
edge states are shown in Fig \ref{edge1} (b). 
 
Phase diagram profile of the system is shown in Fig \ref{topological_phase} 
for $h=1.0$, $\Gamma_z=0$, $\Gamma^\prime_z=0.5$ and when $J^\prime=K^\prime$. 
Three diagrams (a), (b) and (c) are drawn by varying $K$ 
and $J^\prime$ but keeping the values of $J$ fixed at 
0.0, $-0.5$ and $-1.0$, respectively. 
Region I corresponds to $C=\mp 4$. This phase dominates when $J=0$, but begins to diminish  
with the decrease of $J$ and ultimately disappears at $J=-1$. 
Likewise, regions II, III, V and VI corresponds to $\mp 2$, $\mp 1$, 
$\pm 1$ and $\pm 2$, respectively. Region III appears only when $J<0$. 
On the other hand, region IV is a trivial one ($C=0$) and it is 
found to appear in every diagram. 
Another nontrivial phase with $C=\pm 3$ 
that does not appear in Fig \ref{topological_phase}, 
but is found in one of a typical 
situation where $J=-2$, $J'=-1$, $K<0$ and $K'>0$. 
Therefore, in total six distinct TMI phases 
appear in this system. 
     \begin{figure}[b]
   \psfrag{G1}{ $\Gamma_z$}
    \psfrag{G2}{ $\Gamma^\prime_z$}
    \psfrag{A}{$A$}
    \psfrag{B}{$B$}
    \psfrag{L}{$L$}
     \centering
  \includegraphics[width=120pt]{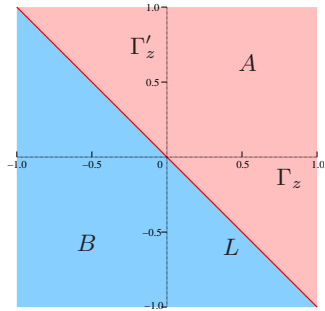}
  \caption{(color online) Topological phase diagram of the system in $\Gamma_z$-$\Gamma^\prime_z$ space.}
  \label{gamma_phase}
   \end{figure}

Interestingly, Chern numbers are found to 
exchange sign with the change of sign of 
$\Gamma_z+\Gamma^\prime_z$, when $h>|\Gamma_z+\Gamma^\prime_z|$. 
This behavior can be understood in terms of TPT 
of the system across the boundary line  $L,\,\Gamma_z+\Gamma^\prime_z=0$, 
which is shown in Fig \ref{gamma_phase}.  
For $J=K=J^\prime=K^\prime=-1$, the system undergoes a transition from phase $\pm 2$ 
in region $A$ to phase $\mp 2$ in region $B$, across the line $L$. 
Similarly, for  $J=K=-1$ and $J^\prime=K^\prime =-0.5$, the system exhibits 
phase $\pm 1$ in region $A$ and $\mp 1$ in region $B$. 
TMI phase does not survive over the line $L$, 
since no band-gap is found there. However, similar kind of 
transition is not observed for the phases  $\pm 3$ and $\mp 4$ for 
obvious reason. 

\begin{figure}[h]
  \psfrag{b}{\tiny $\kappa_{xy} \hbar/k_{B}$}
   \psfrag{a}{\tiny $k_B T$}
 \psfrag{c1}{\tiny (a)}
   \psfrag{c2}{\tiny (b)}
   \psfrag{c3}{\tiny (c)}
   \psfrag{c4}{\tiny (d)}
   \psfrag{c5}{\tiny (e)}
   \psfrag{c6}{\tiny (f)}
    \centering
  \includegraphics[width=200pt]{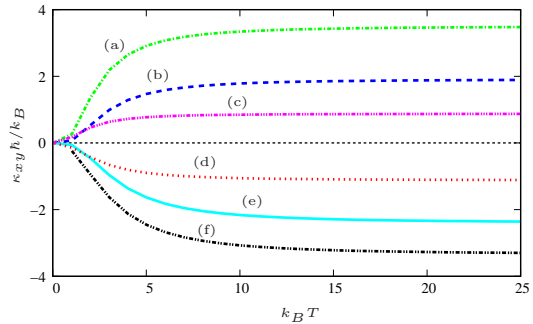}
  \caption{(color online) Variation of $\kappa_{xy}(T)$ with $T$ for 
(a) $J=-0.5$, $J^\prime=-1$, $K^\prime=-1$, for $C=\mp 4$, 
(b) $J=-1$, $J^\prime=-1$, $K^\prime=-1$, for $C=\mp 2$, 
(c) $J=-1$, $K=-1$, for $C=\mp 1$, 
(d) $J=-1$, $K=-1$, $K^\prime=-1$, for $C=\pm 1$, 
(e) $J=-1$, $K=-1$, $J^\prime=-1$, $K^\prime=-1$, for $C=\pm 2$, 
(f) $J=-2$, $K=-0.2$, $J^\prime=-1$, $K^\prime=0.4$ for $C=\pm 3$, 
with $\Gamma^\prime_z=0.5$ and $h=1$. No value is specified for 
those parameters when they are zero.}
   \label{Hall}
  \end{figure}

THC, $\kappa_{xy}$, of the system can be expressed in 
terms of $F(\bold{k})$ for the individual magnon bands as \cite{Murakami1,Murakami2}, 
   \begin{equation}
  \begin{aligned}
   \kappa_{xy}(T)=-\frac{k^2_B T}{4\pi^2\hbar}\,\sum\limits_{n=\pm}\, \iint_{BZ} 
c(\rho_{n}(\bold{k}))\,F_{n}(\bold{k})\,dk_x dk_y. 
  \end{aligned}
\end{equation}
Here $T$ is the temperature, $k_B$ is the Boltzmann constant 
and $\hbar$ is the reduced 
Planck's constant. $F_n(\bold{k})$ is the Berry curvature of the $n$-th band. 
$c(x)=(1+x)\left(\ln{\frac{1+x}{x}}\right)^2-\left(\ln x\right)^2 -2{\rm Li}_2(-x)$, where
${\rm Li}_2(z)=-\int_{0}^{z} du \frac{\ln{(1-u)}}{u}$ 
and $ \rho_{n}(\bold{k})$ is the Bose-Einstein distribution, {\em i.e.},
$\rho_{n}(\bold{k})=1/(e^{E^n_{\bold{k}}/k_{B}T} -1)$. 
$\kappa_{xy}(T)$ gets saturated at high temperatures 
like other thermodynamic quantities.  
The variation of 
$\kappa_{xy}(T)$ with temperature 
will be different for different topological phases since it directly 
depends on the Berry curvatures which actually determine the 
value of Chern numbers as well. 
Variation of $\kappa_{xy}$ with respect to $T$ 
for six different TMI phases is shown in Fig \ref{Hall} with different colors.   
Saturated value of $\kappa_{xy}$ is positive (negative) when the sign of $C$ for 
upper band is negative (positive). On the other hand, absolute value of 
$\kappa_{xy}$ is found proportional to the value of $C$. 

Also, transition among the various 
topological phases can be identified by studying the variations of $\kappa_{xy}$ 
in the parameter space for fixed value of $T$. 
There is a discontinuity in $\kappa_{xy}$ when the system crosses topological phase boundary.
 \begin{figure}[h]
  \psfrag{D1}{\tiny $\kappa_{xy} \hbar/k_{B}$}
   \psfrag{A1}{\tiny $K^\prime$}
   \psfrag{a}{\tiny(a)}
   \psfrag{A}{\tiny $J=-1$}
   \psfrag{B}{\tiny $J^\prime=0$}
   \psfrag{C}{\tiny $K=-1$}
   \psfrag{D}{\tiny $\Gamma_z=0$}
   \psfrag{E}{\tiny $\Gamma^\prime_z=0.5$}
   \psfrag{F}{\tiny $h=1$}
   \psfrag{G}{\tiny $k_B T=20$}
   
     \begin{minipage}{0.21\textwidth}
     \centering
 \includegraphics[width=127pt]{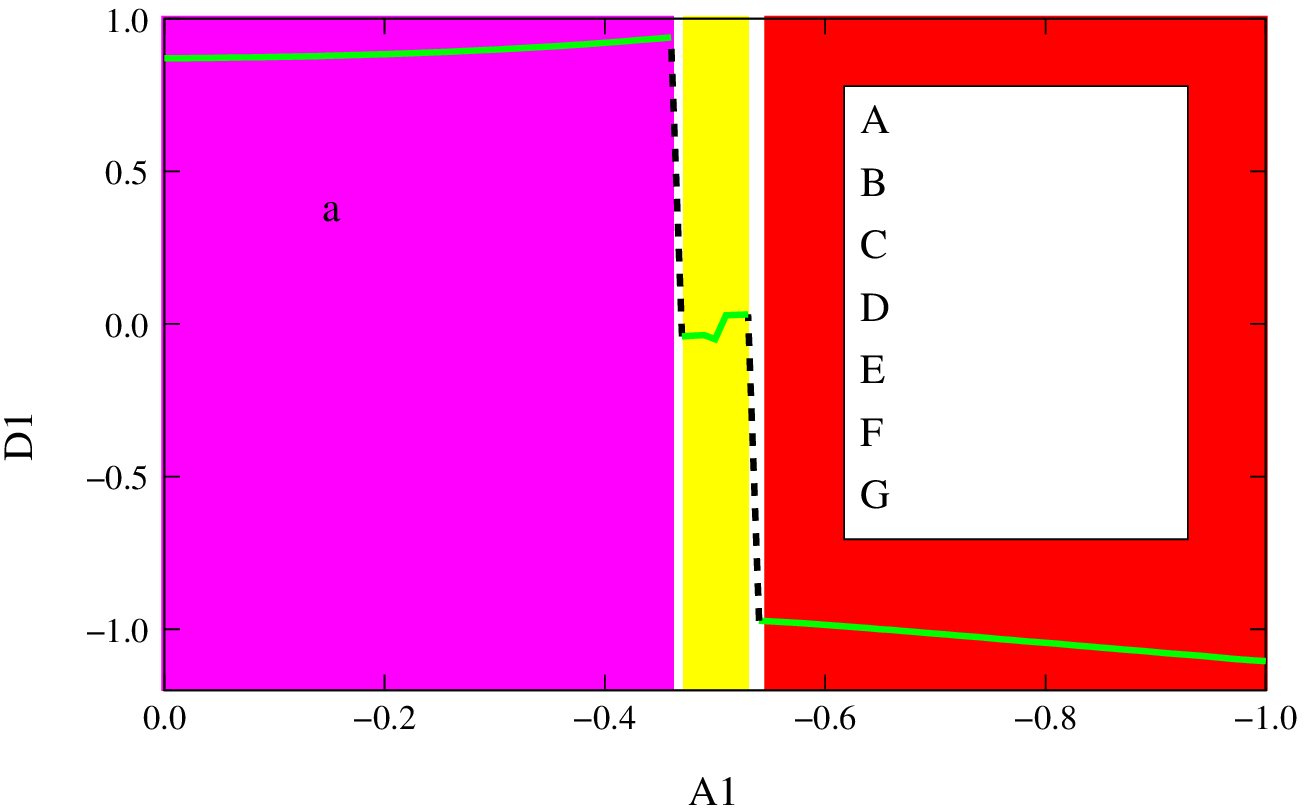}
   \end{minipage}\hfill
    \psfrag{A1}{\tiny $K$}
    \psfrag{b}{\tiny(b)}
     \psfrag{a}{\tiny(a)}
   \psfrag{A}{\tiny $J=-1$}
   \psfrag{B}{\tiny $J^\prime=-1$}
   \psfrag{C}{\tiny $K^\prime=-1$}
   \psfrag{D}{\tiny $\Gamma_z=0$}
   \psfrag{E}{\tiny $\Gamma^\prime_z=0.5$}
   \psfrag{F}{\tiny $h=1$}
   \psfrag{G}{\tiny $k_B T=20$}
    \begin{minipage}{0.235\textwidth}
    \centering
   \includegraphics[width=122pt]{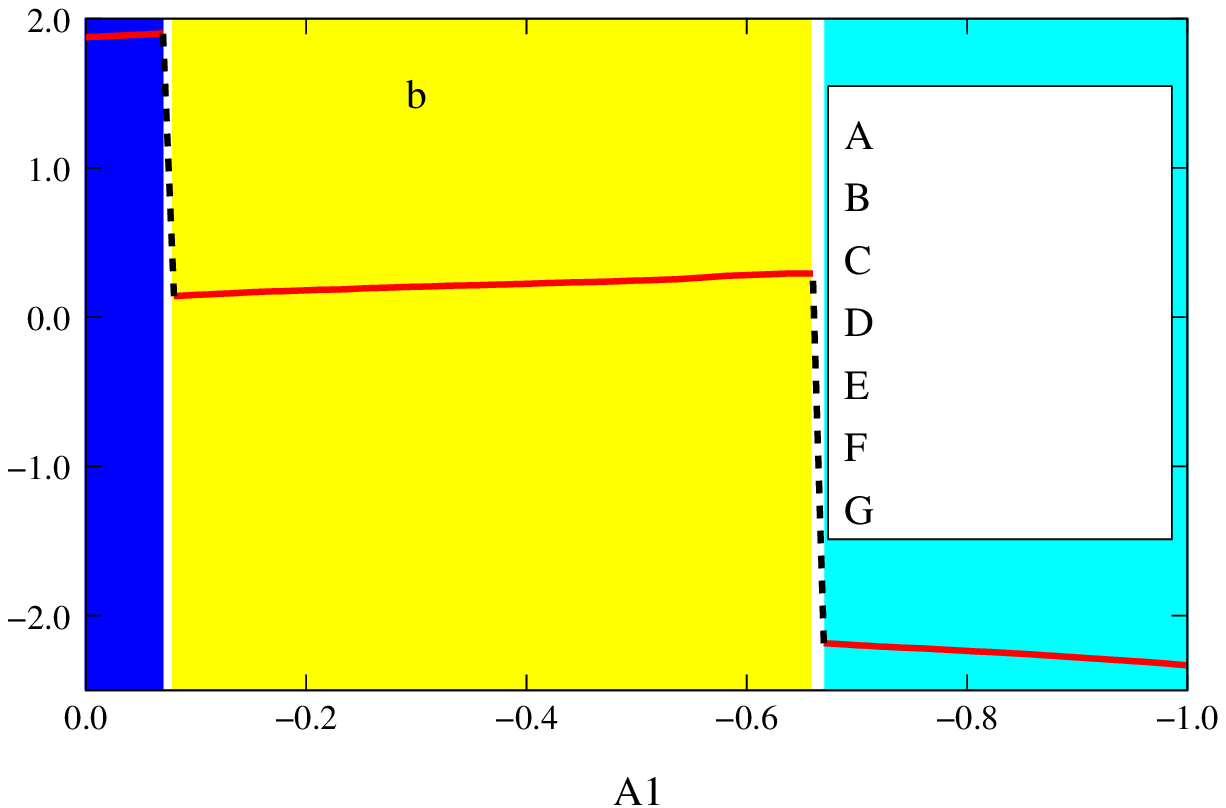}
  \end{minipage}\hfill
  \caption{(color online) Variation of $\kappa_{xy}$ in the parameter space when 
 $T$ is fixed. Different regions are identified with distinct colors, 
those used before in Fig \ref{topological_phase}.}
   \label{Hall1}
  \end{figure}
TPT of the system with respect to 
Kitaev terms $K'$ and $K$ are shown in Figs \ref{Hall1} (a) and (b), respectively. 
In each case, the system undergoes two distinct transitions separating two 
different TMI phases by an intermediate trivial phase $C=0$. 
The system remains trivial at the transition points since no band 
gap is found to appear there. Similar kind of transitions can be observed 
by varying the other parameters as well. For the characterization of TMI phase for  
$\alpha$-RuCl$_3$, values of the parameters in unit of meV 
are given, $J_1=-1.8$, $K_1=-10.6$, 
$J_3=1.25$, $K_3=0.65$ \cite{Gong}. This material is %$\Gamma_1=3.8$, 
topologically identified with the phase $C=\mp 1$. 
Similarly, Li$_2$IrO$_3$ ($J_1=-3.0$, $K_1=-8.0$, $J_3=6.0$) \cite{Valenti}, 
Na$_2$IrO$_3$ ($J_1=3.6$, $K_1=-18.0$, $J_3=1.8$) \cite{Das}, 
and CrI$_3$ ($J_1=-0.2$, $K_1=-5.2$) \cite{Hammel} topologically 
belong to the same phase under external magnetic field.  % $\Gamma_1=9.0$,%$\Gamma_1=2.4$,

\section{Discussion}
\label{discussion}
This investigation reveals the emergence of multiple topological 
phases in Kitaev-Heisenberg ferromagnetic model on honeycomb lattice in 
the presence of both symmetric spin-anisotropic interaction and 
external magnetic field. The NN KHSA model exhibits a single 
topological phase ($C=\pm 1$) when ${\boldsymbol h}$ 
acts along the $z$-axis. Another phase with $C=\mp 1$ 
appears if ${\boldsymbol h}$ points towards [111] direction \cite{Joshi}. 
Here, it has been shown that in total six different topological phases 
($C=\pm 1, \mp 1,\pm 2, \mp 2, \pm 3, \mp 4$) appear with the inclusion of 
NNNN terms when ${\boldsymbol h}$ is directed along the $z$-axis. 
Therefore, a single FM ground state hosts six different TMI phases 
on the basis of magnon excitations. 
Whereas, 
no additional phase emerges when ${\boldsymbol h}$ points towards [111] direction. 
Again in contrast, no additional phase appears in the FM ordered regime 
under any condition with the inclusion of NNN terms. 
FM ground state bears the $U(1)$ rotational symmetry around the 
$z$ direction, although the Hamiltonian, Eq \ref{ham},  
does not comply with this symmetry. 
Analytic expressions of magnon dispersion relations are 
obtained following the LSWT based on this FM ground state. 
Thus, as far as the single magnon mode is concerned, 
the dispersion relations are exact and valid for any value 
of $S$. However, in this analysis, contribution of multi-magnon states is not 
taken into account, since, in a previous study,  
stability of those TMI phases based on the LSWT is found robust 
against the magnon-magnon interactions. 
Contribution of magnon-magnon interactions in the 
KHSA Hamiltonian has been 
treated before (i) perturbatively, 
(ii) nonperturbatively using density matrix 
renormalization group and (iii) numerically using exact 
diagonization of the Hamiltonian on a 24-site cluster \cite{Penc}.  
It has been shown that topologically protected edge states 
obtained in LSWT are robust enough to withstand the 
presence of magnon-magnon interactions. 

Exhaustive topological phase diagrams 
of this model are presented. Six different TMI phases are also 
characterized in terms of thermal behavior of Hall conductivity. 
TMI phase with the highest value of Chern number, $C=\mp 4$ dominates 
when $J=0$, but begins to disappear   
with the decrease of $J$ and ultimately vanishes at $J=-1$,  
when $J'=K'$. On the other hand, TMI phase with $C=\pm 3$ appears 
under various conditions.  
TPT driven by the 
parameters is also studied. Kitaev materials 
are found to belong into a definite TMI phase with $C=\mp 1$. 
Experimental determination of THC for the Kitaev materials 
thus provides a natural way for the verification of their topological phases. 
\section{ACKNOWLEDGMENTS}
MD acknowledges the UGC fellowship, no. 524067 (2014), India. 
AKG acknowledges BRNS-sanctioned 
research project, no. 37(3)/14/16/2015, India.

\end{document}